\documentclass[10pt,aps,prd,twocolumn,showpacs,superscriptaddress,longbibliography,nofootinbib]{revtex4-2}
\usepackage{mathrsfs,amsmath,amsthm,latexsym,amssymb,amsfonts,epsfig,cancel,enumerate,graphicx,txfonts,diagbox}
\usepackage[utf8]{inputenc}  % For UTF-8 encoded text
\usepackage[T1]{fontenc}     % For proper font encoding
\usepackage{lmodern}         % Optional: better fonts for accented characters
\usepackage{multirow}
\usepackage[T1]{fontenc}
\usepackage[utf8]{inputenc}
\usepackage[colorlinks=true,linkcolor=blue,citecolor=blue,urlcolor=blue]{hyperref}
\usepackage{orcidlink}

\usepackage{xcolor}
\usepackage{booktabs}
\usepackage{graphicx}
\usepackage{subcaption}

\definecolor{navy}{RGB}{0,0,150}
\usepackage{appendix}
\allowdisplaybreaks
\newcommand{\RGU}{Department of Physics, The Assam Royal Global University, Guwahati-781035, Assam, India}
\newcommand{\IMS}{The Institute of Mathematical Sciences, C. I. T. Campus, Taramani Chennai, 600113, India}
\newcommand{\ABTU}{Department of Physics, Al-Hussein Bin Talal University, 71111, Ma’an, Jordan}
\begin{document}

\title{Thermodynamics and Optical Properties of Charged Black Holes in Bumblebee gravity Sourced by a Cloud of Strings}

\author{Faizuddin Ahmed\orcidlink{0000-0003-2196-9622}}
\email{faizuddinahmed15@gmail.com}
\affiliation{\RGU}

\author{Shubham Kala\orcidlink{0000-0003-2379-0204}}
\email{shubhamkala871@gmail.com}
\affiliation{\IMS}

\author{Ahmad Al-Badawi\orcidlink{0000-0002-3127-3453}}
\email{ahmadbadawi@ahu.edu.jo}
\affiliation{\ABTU}

\date{\today}

\begin{abstract}
In theories where the Lorentz symmetry of gravity is spontaneously broken, a non-minimally coupled bumblebee vector field acquires a nonzero vacuum expectation value, leading to modifications of standard General Relativity (GR). In this work, we investigate exact solutions describing static and spherically symmetric charged black holes surrounded by a cloud of strings within the framework of bumblebee gravity. We begin by analyzing the thermodynamic properties of these black hole solutions, including their mass, temperature, and entropy, highlighting how Lorentz-violating effects alter standard results. Next, we examine the optical properties of the spacetime, focusing on the photon sphere, the resulting black hole shadow, and the deflection of light, thereby providing potential observational signatures of Lorentz violation. Finally, we explore the impact of Lorentz-violating parameters on classical gravitational tests within the Solar System, such as advance of perihelion precession, in order to set observational constraints. Our analysis provides a comprehensive investigation of the interplay between Lorentz violation, black hole physics, and cloud of strings, offering a framework to probe new physics beyond GR.
\end{abstract}

\maketitle

%\tableofcontents

\section{Introduction}\label{sec:1}

Einstein's general theory of relativity is currently the most successful theory describing gravitation and has been rigorously confirmed through a wide range of experimental and observational tests. Classic confirmations include the perihelion precession of Mercury, the deflection of light by gravity, gravitational redshift, and the Shapiro time delay of radar signals. More recently, the direct detection of gravitational waves from binary black hole mergers \cite{Abbott2016, Abbott2017, Abbott2017GW} and the imaging of the shadows of supermassive black holes, including M87* \cite{EHTL1,EHTL4,EHTL6} and Sgr A* \cite{EHTL12,EHTL16,EHTL17}, have provided further strong evidence supporting the predictions of GR. Despite its remarkable success in explaining a wide range of astrophysical phenomena, certain observational results indicate possible limitations of the theory. For instance, observations of the accelerated expansion of the universe suggest the existence of dark energy, whose physical origin remains unknown within the framework of GR. These unresolved issues motivate the exploration of modifications or extensions of the theory to better understand gravity at cosmological scales.

In the pursuit of a consistent theory of quantum gravity, it is essential to revisit the fundamental symmetries underlying modern physics. In particular, Lorentz symmetry, which forms a cornerstone of both the Standard Model of particle physics and GR, has traditionally been assumed to hold exactly at all energy scales. However, theoretical investigations suggest that at the Planck scale (around $10^{19}$~GeV) this symmetry may be broken \cite{Amelino2013}, implying that Lorentz invariance might only be an approximate symmetry in extreme energy regimes. Complementing these theoretical insights, high-precision experiments and astrophysical observations have indicated potential signatures of Lorentz symmetry violation, such as deviations in photon propagation and gravitational interactions, motivating further scrutiny of this fundamental principle \cite{Friedman2019, Shao2019}. 

To systematically study the consequences of such violations, several theoretical frameworks have been proposed. Among them, the Bumblebee gravity model provides one of the simplest and most widely studied realizations \cite{Kostelecky1989,Kostelecky1989b,Kostelecky2004}. In this model, a vector field acquires a nonzero vacuum expectation value due to a self-interaction potential, spontaneously breaking Lorentz symmetry and inducing anisotropies in the spacetime geometry. This spontaneous symmetry breaking leads to rich phenomenological consequences, particularly in the strong-field regime near compact objects. Noted that spontaneous Lorentz violation arises from a potential \(V(B^\mu B_\mu )\) acting on an vector field \(B_\mu\) \cite{Kostelecky2004,Kostelecky2011}. Not just in Minkowski’s spacetime, the bumblebee model can also be explored in Riemann and Riemann-Cartan spacetimes \cite{Escobar2017,Li2020}. 

Recent work in Bumblebee gravity has revealed a wide variety of exact solutions, illustrating the diversity of possible spacetime structures in the presence of Lorentz violation. These include Schwarzschild-like black holes \cite{Casana2018}, de Sitter–Schwarzschild solutions incorporating a cosmological constant \cite{Maluf2021}, Schwarzschild-like black holes with topological defects \cite{Gullu2022}, solutions in Snyder noncommutative spacetime \cite{Jha2021}, new Schwarzschild-like and Schwarzschild-(A)dS-like black hole solutions \cite{LiuWu2025}, static spherical vacuum solutions \cite{Xu2023} and with general VEVs \cite{ZhuLi2025}, Schwarzschild-like black holes in metric-affine bumblebee gravity \cite{ Lambiase2023, Araujo2024}. In addition, exotic geometries such as traversable wormholes \cite{Ovgun2019,Filho2025} and charged black holes \cite{Li2026} have also been explored. Collectively, these studies demonstrate that Lorentz-violating effects can lead to significant modifications of classical gravitational structures, providing a concrete arena to test quantum gravity phenomenology and search for Planck-scale deviations from GR.

Moreover, rotating black hole solutions have also been extensively explored within the framework of Bumblebee gravity. Notable examples include Kerr-like solutions \cite{Li2020,DingLiu2020}, charged black holes \cite{Liu2025}, Taub-NUT–like black holes \cite{ChenLiu2025}, BTZ-like black holes \cite{Ding2023}, Kerr-Sen-like solutions \cite{Jha2021b,Poulis2022}, G\"{o}del-like solution \cite{Santos2015}, and other \cite{ Mangut2023, Uniyal2023,Pantig:2024lpg}, highlighting the rich structure of Lorentz-violating rotating space-times within this theory.

Black hole thermodynamics plays a central role in the search for a consistent theory of quantum gravity, providing a deep link between GR, quantum mechanics, and statistical physics. The thermodynamic interpretation of black holes was first proposed in Refs.~\cite{Bekestein1972}, where it was stated that if black holes had zero entropy, the total entropy of the Universe would decrease when matter carrying entropy falls into them, thereby violating the second law of thermodynamics. To resolve this inconsistency, the author in \cite{Bekestein1973} proposed that black holes themselves must possess an entropy $S_\mathrm{BH}$ proportional to the area $A$ of their event horizon: \(S_\mathrm{BH} = \eta \, k_B\,A/\ell_\mathrm{P}^2,\) where $\ell_\mathrm{P} = \sqrt{\hbar G/c^3}$ is the Planck length, $k_B$ is the Boltzmann constant, and $\eta$ is a dimensionless proportionality constant of order unity, later fixed to $\eta = 1/4$ by Hawking's semiclassical analysis~\cite{Hawking1975}. Shortly thereafter, Hawking~\cite{Hawking1975} demonstrated that black holes are not completely black but emit thermal radiation due to quantum effects near the event horizon. The temperature of this radiation, now known as the Hawking temperature, is given by \( T_\mathrm{H} = \frac{\hbar \kappa}{2 \pi k_B c},\) where $\kappa$ is the surface gravity of the black hole. This discovery implies that black holes behave as thermodynamic objects with a well-defined temperature and entropy. Furthermore, Hawking showed that in any classical process, the surface area $A$ of a black hole horizon cannot decrease~\cite{Hawking1975}, a result formally analogous to the second law of thermodynamics: \(\delta A \geq 0.\) These insights culminated in the Bekenstein-Hawking area law, which establishes that the entropy of a black hole scales with the area of its horizon rather than its volume: \(S_\mathrm{BH} = \frac{k_B c^3 A}{4 \hbar G}.\) This remarkable connection suggests that the fundamental degrees of freedom of gravity may be encoded holographically on the two-dimensional surface of the horizon, laying the foundation for subsequent developments in quantum gravity, including the holographic principle and the study of black hole microstates in string theory and other quantum gravity frameworks~\cite{Bardeen1973}. The thermodynamic properties of black hole solutions in Einstein‑bumblebee gravity, both in asymptotically flat and AdS backgrounds, have been extensively studied \cite{MaiXuLiang2023, HFDing2025, YuSen2024, Kaur2025, ChenLiuPRD2025, Riasat2025, DingShi2023,Kanzi2019}.

Letelier~\cite{PSL1979} first introduced the concept of a cloud of strings, which can be viewed as a one-dimensional analogue of a cloud of dust. A black hole surrounded by such a string cloud can be interpreted as a purely energetic field configuration, where the radial distribution of strings generates an effective negative pressure that counteracts the inward gravitational pull, analogous to the effect of a global monopole~\cite{Barriola1989}. In four-dimensional spacetimes, this results in a modification of the Schwarzschild solution, with the string cloud acting as an additional source of gravity that alters the spacetime geometry~\cite{Dey2018}. Notably, the presence of a string cloud typically increases the radius of the event horizon relative to the standard Schwarzschild black hole. Solving Einstein’s field equations with a string cloud provides a useful phenomenological model for relativistic string distributions in gravitational contexts. A key advantage of the string cloud framework is that it can be naturally generalized to higher-dimensional spacetimes, making it versatile for studies in both Einstein gravity and various modified gravity theories. Consequently, black hole solutions incorporating clouds of strings have been extensively investigated in the literature~\cite{FA1,FA2,FA3,FA4,CS1,CS2,CS3,CS4,CS5,CS6,CS7,Vishvakarma:2024icz,Kala:2024fvg}.

The aim of this manuscript is to investigate charged black holes surrounded by a cloud of strings within Einstein’s field equations in the presence of Lorentz symmetry violation. We analyze how both the string cloud and spontaneous Lorentz symmetry breaking affect the thermodynamic properties of the black hole. Furthermore, we study the impact of these effects on the optical properties, including the black hole shadow, photon sphere, and the deflection of light. The dynamics of massive test particles are also examined, with a focus on particle trajectories and, in particular, the perihelion precession. Finally, we investigate the rate of energy emission from the black hole, demonstrating how the presence of the string cloud and Lorentz-violating effects modify this emission rate.

The paper is organized as follows. In Section \ref{sec:2}, we introduce the spacetime metric and set up the background geometry for the black hole under consideration. Section \ref{sec:3} discusses the thermodynamic properties of the system, while Section \ref{sec:4} examines its optical features, including photon spheres and shadows. In Section \ref{sec:5}, we analyze the impact of in-falling accretion gas on the black hole shadow, followed by Section \ref{sec:6}, which presents solar system tests to constrain the model parameters. Section \ref{sec:7} focuses on the bending of light in the black hole spacetime, and Section \ref{sec:8} investigates the rate of energy emission. Section \ref{sec:9} addresses the sparsity of Hawking radiation, highlighting quantum and classical effects, and finally, Section \ref{sec:10} concludes the study by summarizing the main findings and potential directions for future research. The system of units are chosen as $c=1=G=\hbar$.

\section{Spacetime metric: Background Set Up}\label{sec:2}

As discussed in the previous section, the Bumblebee model extends GR by introducing a vector field, known as the Bumblebee field, which couples nonminimally to gravity. The vector field \(B_\mu\) acquires a nonzero vacuum expectation value via a suitable potential, leading to the spontaneous breaking of Lorentz symmetry in the gravitational sector~\cite{Liu2025}. In this work, we incorporate a cloud of strings into a charged, spherically symmetric black hole solution within Bumblebee gravity, as derived in~\cite{Liu2025}. The line-element describing a charged spherically symmetric black hole solutions, given by \cite{Liu2025,Li2026}:
\begin{equation}
\mathrm{d} s^2=-h(r)dt^{2}+\frac{1+\ell}{h(r)}dr^{2}+r^{2}(d\theta ^{2}+\sin^{2} \phi d\phi^2),,\label{aa1}
\end{equation}
where,
\begin{align}
h(r)=1-\frac{2M}{r}+\frac{2(1+\ell)\,q^{2}}{(2+\ell)\,r^{2}}, \label{aa2}
\end{align}
where $\ell=\xi \bar{b}^2$ is Lorentz-violating (LV) parameter and \(q\) is the charge of the black hole.

To include string-like objects, we consider Nambu-Goto action given by \cite{PSL1979}
\begin{equation}
    S_{\rm CS}=\int \sqrt{-\gamma}\,\mathcal{M}\,d\lambda^0\,d\lambda^1=\int \mathcal{M}\sqrt{-\frac{1}{2}\,\Sigma^{\mu \nu}\,\Sigma_{\mu\nu}}\,d\lambda^0\,d\lambda^1,\label{aa3}
\end{equation}
where $\mathcal{M}$ is the dimensionless constant which characterizes the string, ($\lambda^0\,\lambda^1$) are the time
like and spacelike coordinate parameters, respectively \cite{JLS1960}. $\gamma$  is the determinant of the induced metric of the strings world sheet given by $\gamma=g^{\mu\nu}\frac{ \partial x^\mu}{\partial \lambda^a}\frac{ \partial x^\nu}{\partial \lambda^b}$.  $\Sigma_{\mu\nu}=\epsilon^{ab}\frac{ \partial x^\mu}{\partial \lambda^a}\frac{ \partial x^\nu}{\partial \lambda^b}$ is bivector related to string world sheet, where $\epsilon^{ab}$ is the second rank Levi-Civita tensor which takes the non-zero values as $\epsilon^{01} = -\epsilon^{10} = 1$.

\begin{equation}
   T_{\mu\nu}^{\rm CS}=2 \frac{\partial}{\partial g_{\mu \nu}}\mathcal{M}\sqrt{-\frac{1}{2}\Sigma^{\mu \nu}\,\Sigma_{\mu\nu}} =\frac{\rho^{\rm CS} \,\Sigma_{\alpha\nu}\, \,\Sigma_{\mu}^\alpha }{\sqrt{-\gamma}}, \label{aa4}
 \end{equation}
where $\rho^{\rm CS}$ is the proper density of the string cloud. The energy-momentum tensor components are given by
\begin{equation}
    T^{t\,(\rm CS)}_{t}=\rho^{\rm CS}=\frac{\alpha}{r^2}=T^{r\,(\rm CS)}_{r},\quad T^{\theta\,(\rm CS)}_{\theta}=T^{\phi\,(\rm CS)}_{\phi}= 0,\label{aa5}
\end{equation}
where $\alpha$ is a constant associated with string-like objects called string parameter.

The presence of a cloud of strings modifies the black hole spacetime by increasing the radius of the event horizon and introducing a conical deficit, analogous to the gravitational effects observed in the spacetime of a global monopole \cite{Barriola1989}. These modifications can have significant astrophysical consequences, influencing particle dynamics, accretion processes, and observational signatures, thereby motivating further investigation into the role and impact of string clouds in strong gravitational fields.

\begin{figure*}[tbhp]
	\centerline{
		\includegraphics[width=80mm,height=70mm]{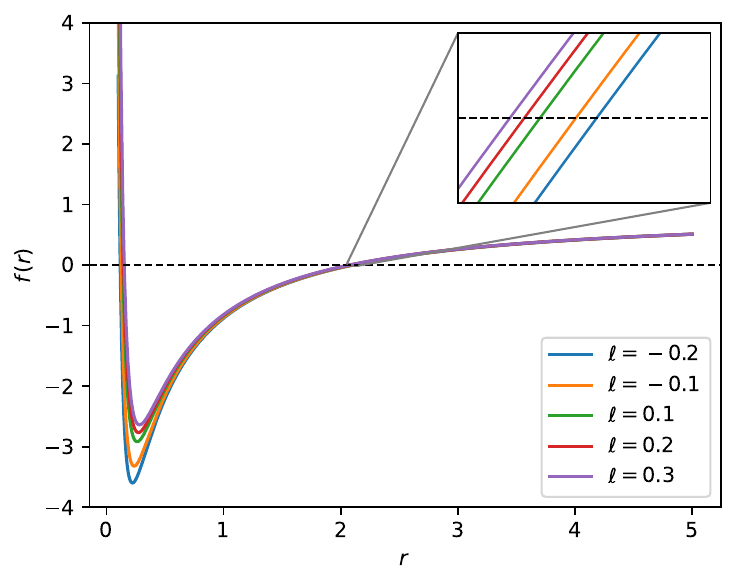}\qquad
        \includegraphics[width=80mm,height=70mm]{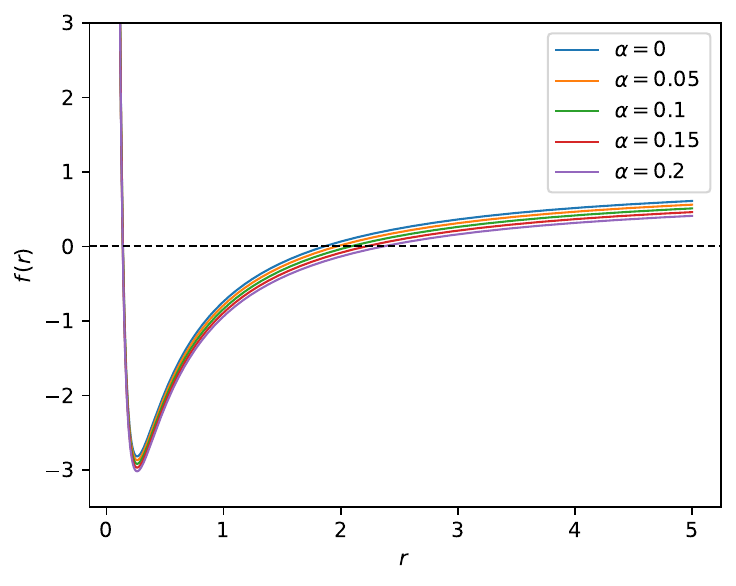}}(a) $\alpha=0.1$ \hspace{5cm} (b) $\ell=0.2$
	\caption{The variation of metric function $f(r)$ with radial distance $r$ for different value of $\ell$ and $\alpha$. The fixed parameters are $q=0.5$, and $M=1$.}
\label{fig:metric-function}
\end{figure*}

Based on the above equations, a charged black hole solution in the bumblebee gravity frame surrounded by a cloud of strings can be expressed as
\begin{equation}
\mathrm{d} s^2=-f(r)dt^{2}+\frac{1+\ell}{f(r)}dr^{2}+r^{2}(d\theta ^{2}+\sin^{2} \phi d\phi^2),\label{metric}
\end{equation}
where,
\begin{align}
f(r)=1-\alpha-\frac{2M}{r}+\lambda\,\frac{q^{2}}{r^{2}},\quad \lambda=\frac{1+\ell}{1+\ell/2}. \label{function}
\end{align}
For the metric described above, if the LV parameter is zero, $\ell=0$, the space-time reduces to the Reissner-Nordström black hole solution with a cloud of strings. When both the LV parameter and charge parameter vanish (i.e., $\ell=0$ and $q=0$), the structure degenerates into the Letelier solution \cite{PSL1979}. Furthermore, when $q=0$ and $\alpha=0$, the solution reduces to the Schwarzschild-like black hole in a bumblebee gravity derived in \cite{Casana2018}. In Fig.~\ref{fig:metric-function}, we plot the metric function as a function of the radial distance. For the chosen values of the parameters $\ell$ and $\alpha$, the metric function exhibits two horizons, corresponding to the black hole (event) horizon and the Cauchy horizon.

The horizons structure can be determined using the condition $f(r_h)=0$ and is given by
\begin{equation}
r_{\pm}=\frac{M}{1-\alpha}\,\left[1 \pm \sqrt{1-\frac{\lambda\,q^2 (1-\alpha)}{M^2}}\right].\label{aa6}
\end{equation}
The horizons exist provided we have the following constraint on the electric charge
\begin{equation}
    q^2 < \frac{M^2}{\lambda\,(1-\alpha)}.\label{aa7}
\end{equation}

The horizons satisfy the following relations:
\begin{align}
    r_{+}+r_{-}&=2M/(1-\alpha),\quad r_{+}\,r_{-}=\lambda\,q^2/(1-\alpha)\nonumber\\
    r_{+}-r_{-}&=\frac{2\sqrt{M^2-\lambda\,q^2\,(1-\alpha)}}{1-\alpha}.\label{aa8}
\end{align}

\begin{figure*}[tbhp]
	\centerline{
		\includegraphics[width=80mm,height=70mm]{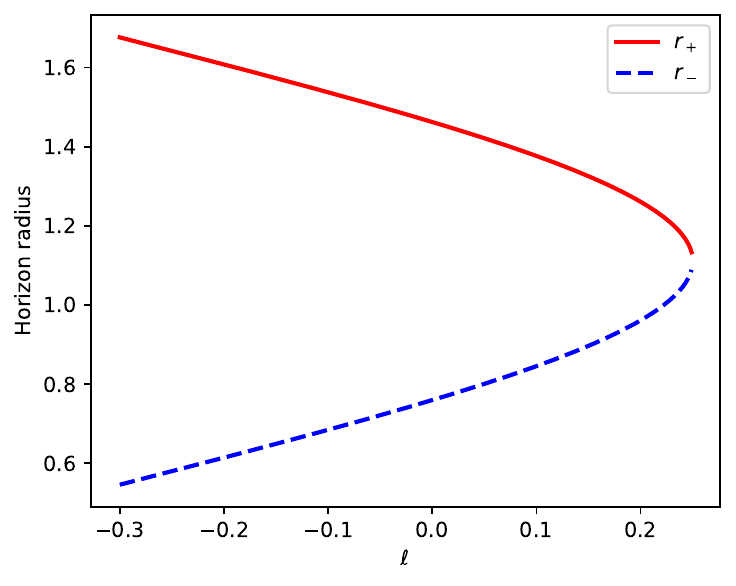}
    \includegraphics[width=80mm,height=70mm]{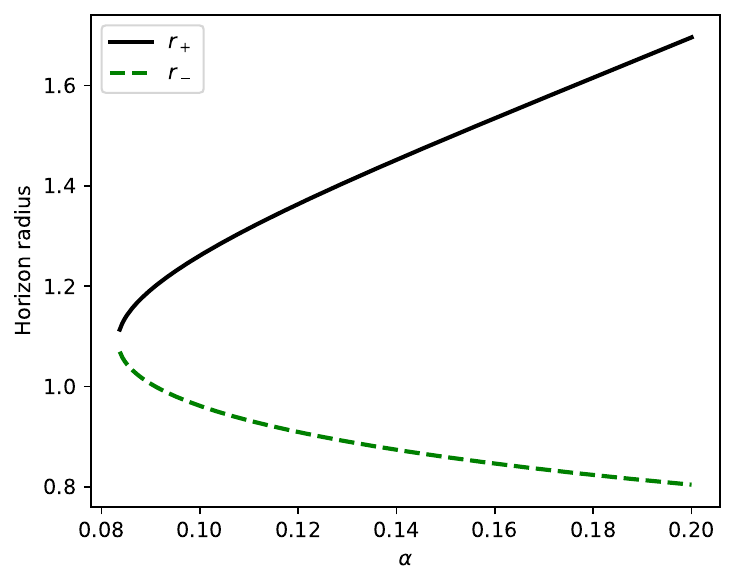}}(a) $\alpha=0.1$ \hspace{5cm} (b) $\ell=0.2$
	\caption{Plots showing the dependence of the black hole event horizon $r_{+}$ and the Cauchy horizon $r_{-}$ on the LV parameter $\ell$ and the CoS parameter $\alpha$. Here, we consider $\ell=0.2$, $\alpha=0.1$ and $q=0.8$.}
\label{figHorizonBHCauchy}
\end{figure*}

Fig.~\ref{figHorizonBHCauchy} depicts the variation of the event horizon $r_{+}$ and the Cauchy horizon $r_{-}$ with respect to the LV parameter $\ell$ (left panel) and the CoS $\alpha$ (right panel), while keeping the remaining parameters fixed. In the left panel, it is observed that the event horizon radius $r_{+}$ decreases monotonically as the bumblebee parameter $\ell$ increases. In contrast, the Cauchy horizon radius $r_{-}$ exhibits an increasing trend with $\ell$. This behavior indicates that the Lorentz symmetry breaking effects encoded in the bumblebee parameter tend to reduce the size of the outer horizon while enlarging the inner horizon. Consequently, the separation between the two horizons diminishes for larger values of $\ell$, suggesting a significant modification of the causal structure of the black hole spacetime.
In the right panel, the dependence of the horizons on the CoS parameter $\alpha$ shows an opposite trend. The event horizon $r_{+}$ increases steadily with increasing $\alpha$, whereas the Cauchy horizon $r_{-}$ decreases. This demonstrates that the CoS correction strengthens the outer horizon while suppressing the inner horizon. Hence, the horizon gap widens with increasing $\alpha$, highlighting the competing roles of the bumblebee parameter and the cosmological parameter in determining the horizon structure. In summary, the analysis shows that the parameters $\ell$ and $\alpha$ exert opposite and complementary influences on the radii of the event and Cauchy horizons. These results emphasize the sensitivity of the horizon geometry to Lorentz-violating effects and cosmological corrections.

\begin{figure*}[tbhp]
	\centerline{
		\includegraphics[width=120mm,height=80mm]{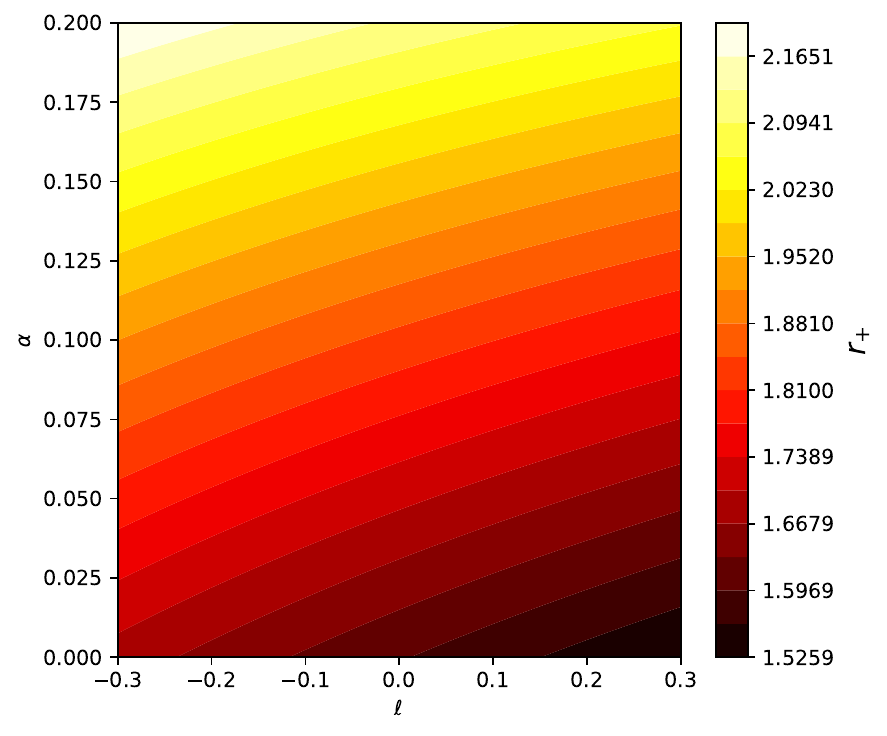}
        }
	\caption{Two--dimensional contour plot showing the variation of black hole horizon with $\ell$ and $\alpha$. Here we fix, $M=1$ and $q=0.8$.}
\label{figHorizon2D}
\end{figure*}

Fig.~\ref{figHorizon2D} presents a two--dimensional contour plot of the event horizon radius $r_{+}$ as a function of the LV parameter $\ell$ and the cosmological parameter $\alpha$, for fixed values of the mass $M=1$ and charge $q=0.8$. The color bar represents the magnitude of the event horizon radius, where lighter shades correspond to larger values of $r_{+}$ and darker shades indicate smaller values. It is evident from the figure that the event horizon radius increases with increasing $\alpha$ and decreases with increasing $\ell$. This behavior is consistent with the trends observed in Fig.~\ref{figHorizonBHCauchy} and confirms the combined influence of both parameters on the horizon structure. Furthermore, the smooth and continuous contour lines indicate that the variation of $r_{+}$ with respect to $\ell$ and $\alpha$ is regular throughout the considered parameter space, with no abrupt transitions or discontinuities. This suggests that the black hole solution remains well defined within the explored ranges of the parameters. The contour plot clearly illustrates the interplay between Lorentz symmetry breaking effects, governed by the LV parameter $\ell$, and CoS corrections, characterized by the parameter $\alpha$, in determining the size of the black hole event horizon.

\section{Thermodynamic Properties of BH }\label{sec:3}

In this section, we analyze the thermodynamic properties of a charged black hole surrounded by a cloud of strings in Bumblebee gravity, focusing on how both the string cloud and Lorentz-violation affect the Hawking temperature. Furthermore, we derive the corresponding first law of black hole thermodynamics and establish the associated Smarr formula, providing a consistent description of the black hole’s energetic and geometric properties.

To study the thermodynamic properties, we re-write the metric (\ref{metric}) as
\begin{equation}
\mathrm{d} s^2=g_{tt}\,dt^{2}+g_{rr}\,dr^{2}+r^{2} d\Omega^2,\label{bb1}
\end{equation}
where,
\begin{align}
g_{tt}=-f(r),\quad g_{rr}=(1+\ell)/f(r) \label{bb2}
\end{align}

The surface gravity is defined by
\begin{equation}
    \kappa=-\frac{1}{2}\,\frac{1}{\sqrt{-g_{tt}\,g_{rr}}}\,\frac{dg_{tt}}{dr}\Big{|}_{r=r_h},\label{bb3}
\end{equation}
Therefore, the Hawking temperature is given by
\begin{equation}
    T=\frac{\kappa}{2\pi}=\,\frac{1}{2 \pi r_h} \,\frac{\frac{M}{r_h}-\lambda\,\frac{q^2}{r_h^2}}{\sqrt{1+\ell}},\label{bb4}
\end{equation}
where the horizon $r_h$ using Eq. (\ref{aa6}) is given by
\begin{equation}
r_h=\frac{M}{1-\alpha}\,\left[1 +\sqrt{1-\frac{\lambda\,q^2 (1-\alpha)}{M^2}}\right].\label{horizon}
\end{equation}

In the limit $q=0$, corresponding to the absence of electric charge, the Hawking temperature simplifies as (now the horizon radius $r_h=2M/(1-\alpha)$)
\begin{equation}
    T=\frac{M}{2 \pi r_h^2} \,\frac{1}{\sqrt{1+\ell}}=\frac{1}{8\pi M}\,\frac{(1-\alpha)^2}{\sqrt{1+\ell}}.\label{bb5}
\end{equation}

\begin{figure*}[tbhp]
	\centerline{
		\includegraphics[width=80mm,height=70mm]{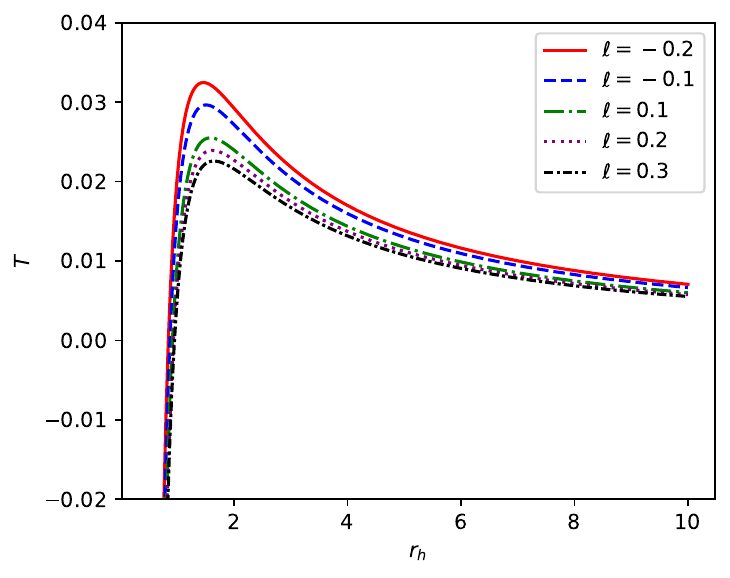}\qquad
        \includegraphics[width=80mm,height=70mm]{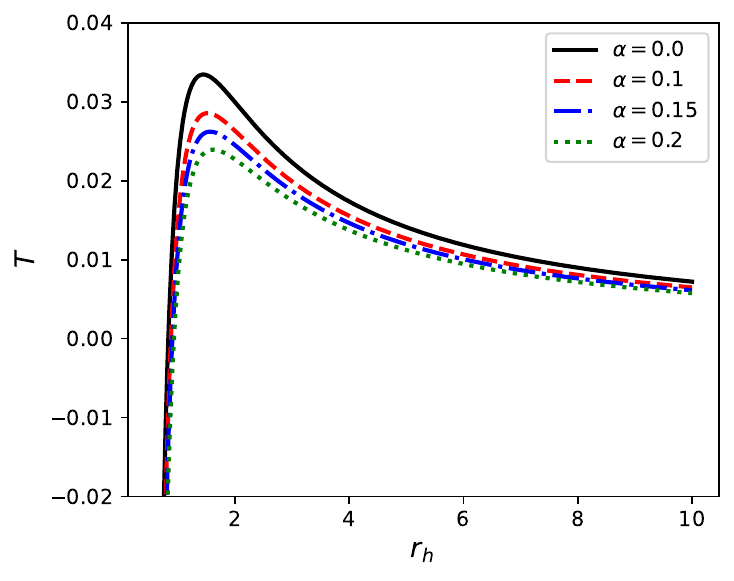}}(a) $\alpha=0.1$ \hspace{5cm} (b) $\ell=0.2$
	\caption{The variation of Hawking temperature as a function of $r_{h}$; for different values of LV parameter ($\ell$) and CoS parameter ($\alpha$). Here we fix, $M=1$, $q=0.8$, $\ell=0.2$ and $\alpha=0.1$ where applicable.}
\label{fig1HT}
\end{figure*}

Fig.~\ref{fig1HT} depicts the variation of the Hawking temperature $T$ as a function of the event horizon radius $r_h$ for different values of the LV parameter $\ell$ and the cosmological parameter $\alpha$, with fixed values of $M=1$ and $q=0.8$. The left panel shows the dependence of $T$ on $r_h$ for several values of $\ell$, while the right panel illustrates the effect of $\alpha$. From both panels, it is evident that the Hawking temperature grows first to a peak value at a critical horizon radius and  then decreases to zero at a constant vale of LV and CoS parameter. For a fixed $r_h$, the temperature attains higher values for smaller (more negative) values of the LV parameter $\ell$, whereas increasing $\ell$ leads to a suppression of the Hawking temperature. This demonstrates that Lorentz symmetry breaking effects reduce the surface gravity and hence lower the thermal radiation rate of the black hole. Similarly, the right panel shows that variations in the CoS parameter $\alpha$ also affect the temperature profile. The Hawking temperature attains maximum values for the case $\alpha=0$ and then decreases with increasing CoS parameter similar to LV parameter. Thus, the obtained results suggest that both Lorentz-violating effects and presence of CoS act to suppress the Hawking temperature, thereby modifying the evaporation characteristics of the black hole.

Considering the Beketsein-Hawking area law of entropy, the systems entropy is given by
\begin{equation}
    S=\mathcal{A}/4=\pi r^2_h.\label{bb6}
\end{equation}
The black hole mass in terms of horizon using the condition $f(r_h)=0$ is given by
\begin{equation}
    M=\frac{r_h}{2}\left(1-\alpha+\lambda\,\frac{q^{2}}{r^{2}_h}\right)=\frac{1}{2}\sqrt{\frac{S}{\pi}}\left(1-\alpha+\lambda\,\frac{\pi\,q^{2}}{S}\right).\label{bb7}
\end{equation}

\begin{figure*}[tbhp]
	\centerline{
		\includegraphics[width=80mm,height=70mm]{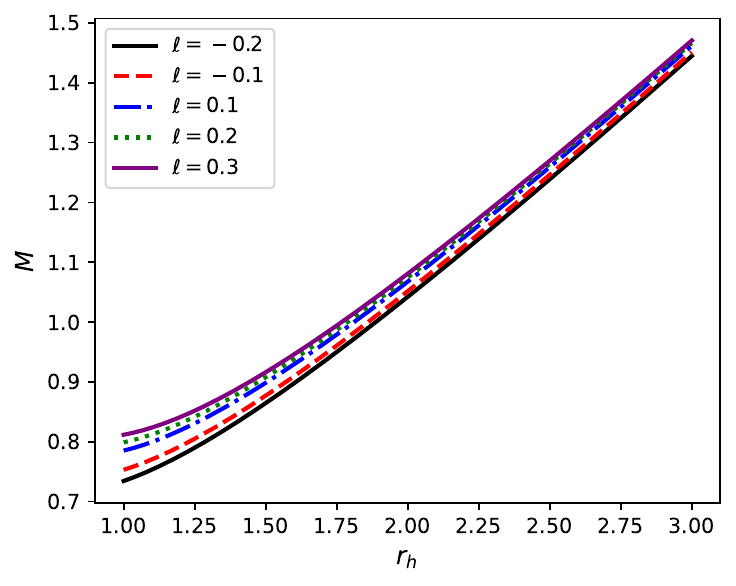}\qquad
        \includegraphics[width=80mm,height=70mm]{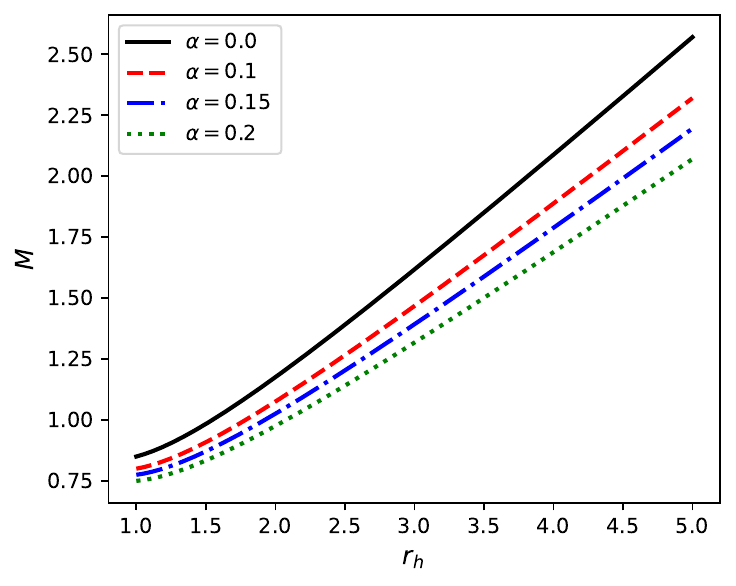}}(a) $\alpha=0.1$ \hspace{5cm} (b) $\ell=0.2$
	\caption{The variation of black hole mass as a function of $r_{h}$; for different values of LV parameter ($\ell$) and CoS parameter ($\alpha$). Here we fix, $M=1$, $q=0.8$, $\ell=0.2$ and $\alpha=0.1$ where applicable.}
\label{fig1HTM}
\end{figure*}

\begin{figure*}[tbhp]
	\centerline{
		\includegraphics[width=80mm,height=70mm]{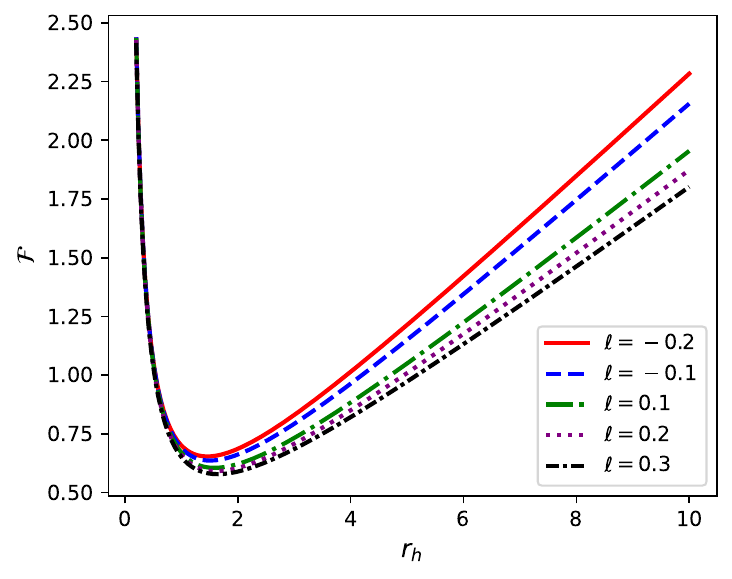}\qquad
        \includegraphics[width=80mm,height=70mm]{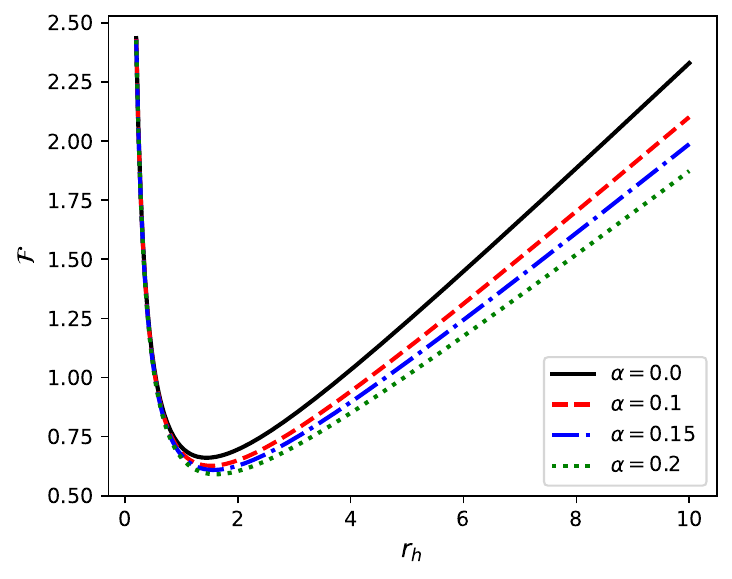}}(a) $\alpha=0.1$ \hspace{5cm} (b) $\ell=0.2$
	\caption{The variation of Helmholtz free energy as a function of $r_{h}$; for different values of LV parameter ($\ell$) and CoS parameter ($\alpha$). Here we fix, $M=1$, $q=0.8$, $\ell=0.2$ and $\alpha=0.1$ where applicable.}
\label{fig1Helmholtz}
\end{figure*}

\begin{figure*}[tbhp]
	\centerline{
		\includegraphics[width=80mm,height=70mm]{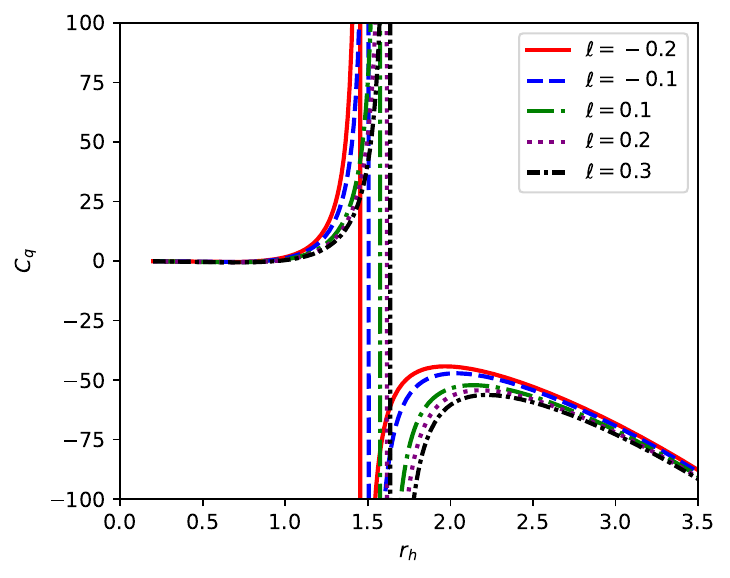}\qquad
        \includegraphics[width=80mm,height=70mm]{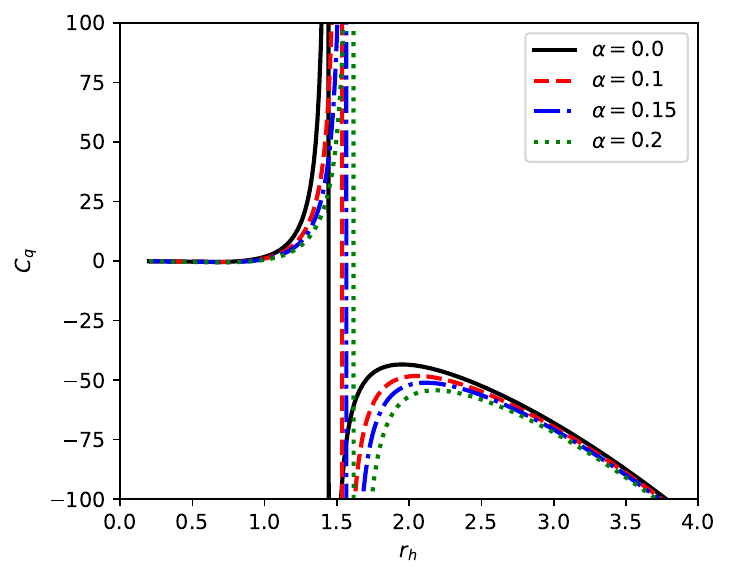}}(a) $\alpha=0.1$ \hspace{5cm} (b) $\ell=0.2$
	\caption{The variation of specific heat as a function of $r_{h}$; for different values of LV parameter ($\ell$) and CoS parameter ($\alpha$). Here we fix, $M=1$, $q=0.8$, $\ell=0.2$ and $\alpha=0.1$ where applicable.}
\label{fig1SpecificHeat}
\end{figure*}

\begin{figure*}[tbhp]
	\centerline{
		\includegraphics[width=80mm,height=70mm]{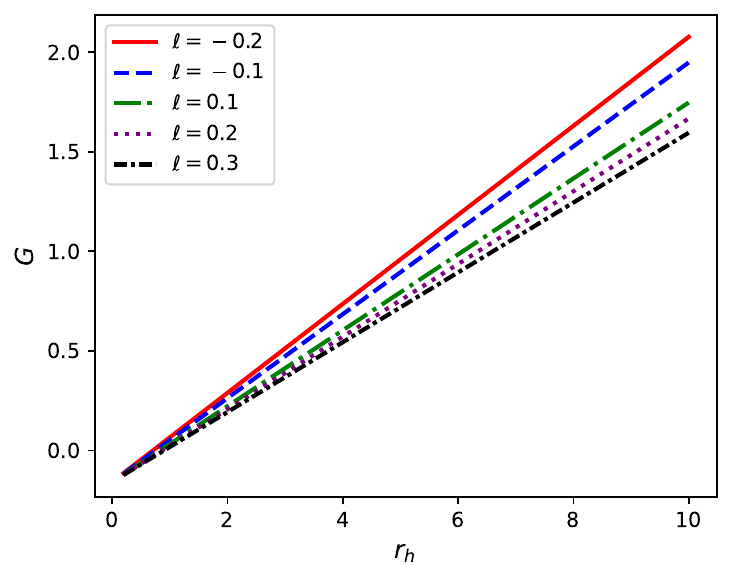}\qquad
        \includegraphics[width=80mm,height=70mm]{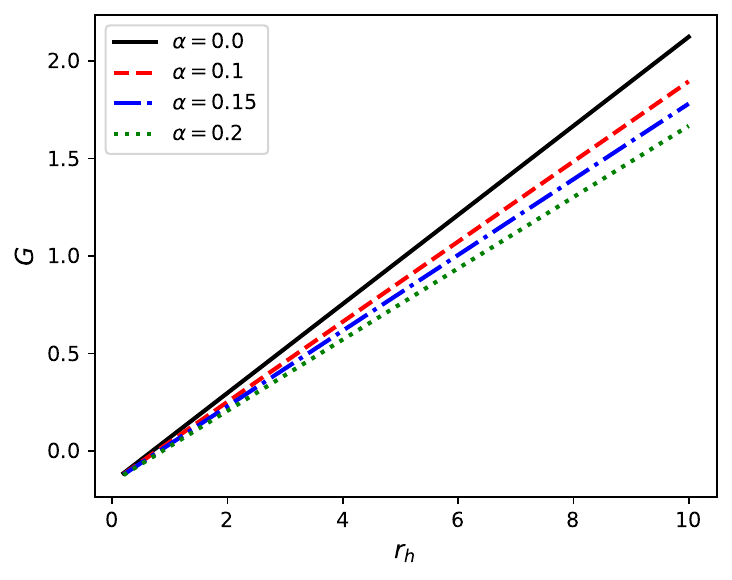}}(a) $\alpha=0.1$ \hspace{5cm} (b) $\ell=0.2$
	\caption{The variation of Gibbs free-energy as a function of $r_{h}$; for different values of LV parameter ($\ell$) and CoS parameter ($\alpha$). Here we fix, $M=1$, $q=0.8$, $\ell=0.2$ and $\alpha=0.1$ where applicable.}
\label{fig1Gibbs}
\end{figure*}

\begin{figure*}[tbhp]
	\centerline{
		\includegraphics[width=180mm,height=140mm]{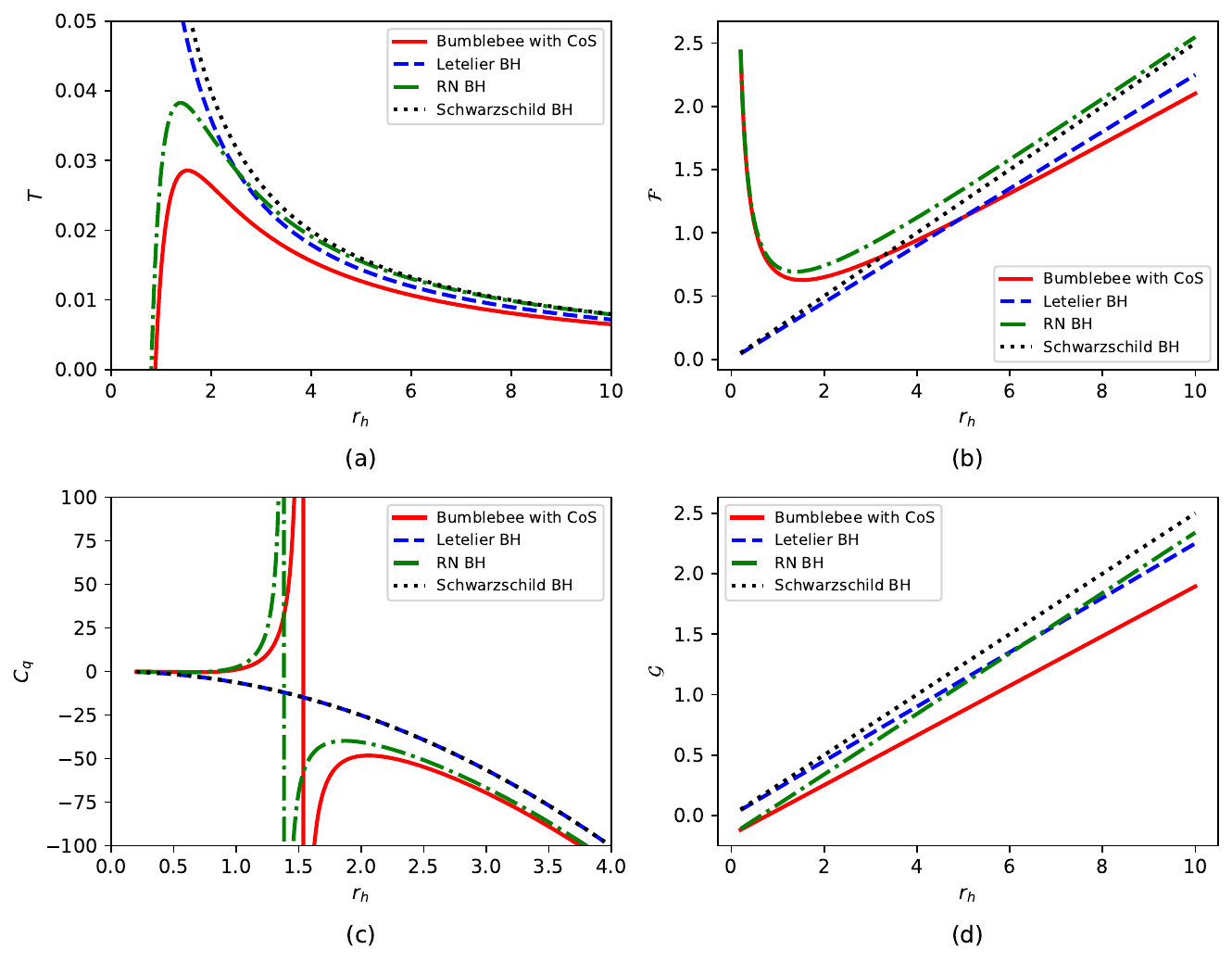}}
	\caption{Comparison of the thermodynamic behavior of the charged black hole in Bumblebee
gravity sourced by a CoS with other well-known cases to which it reduces in specific limits. Here we fix, $M=1$, $q=0.8$, $\ell=0.2$ and $\alpha=0.1$ where applicable.}
\label{ThermodynamicsComprision}
\end{figure*}

Fig.~\ref{fig1HTM} illustrates the variation of the black hole mass $M$ as a function of the horizon radius $r_h$ for different values of the LV parameter $\ell$ and the cosmological parameter $\alpha$, with fixed values of $M=1$ and $q=0.8$. The left panel shows the influence of $\ell$, while the right panel demonstrates the effect of $\alpha$ on the mass-radius relation. In both panels, the black hole mass increases monotonically with the horizon radius, confirming the expected physical behavior that larger event horizons correspond to more massive black holes. For a fixed horizon radius, the mass is found to increase with increasing $\ell$, indicating that Lorentz-violating corrections enhance the effective gravitational mass of the black hole. This suggests that the LV parameter contributes positively to the energy content of the spacetime. On the other hand, the right panel shows that the mass decreases systematically with increasing $\alpha$. This behavior implies that the cosmological correction associated with the CoS parameter weakens the gravitational field, leading to a reduction in the mass required to sustain a given horizon radius. The smooth and continuous nature of the curves in both panels indicates the absence of any critical behavior or phase transition within the explored parameter ranges. These results demonstrate that both the LV parameter and the cosmological parameter play significant roles in modifying the mass-radius relation and, consequently, the thermodynamic structure of the black hole.

Therefore, the ADM mass can be determined using the the Abbott-Deser (AD) method \cite{AbbottDeser1982} and is given by
\begin{equation}
    \mathcal{M}=\frac{M}{\sqrt{1+\ell}}.\label{bb8}
\end{equation}

From the above, it is clear that the ADM mass $\mathcal{M}=\mathcal{M}(S, q)$ and hence, the differential mass can be expressed as,
\begin{equation}
    d\mathcal{M}=\left(\frac{\partial \mathcal{M}}{\partial S}\right)_{q}\,dS+\left(\frac{\partial \mathcal{M}}{\partial q}\right)_{S}\,dq=T\,dS+\Phi\,dq,\label{bb8a}
\end{equation}
where
\begin{align}
    T&=\left(\frac{\partial \mathcal{M}}{\partial S}\right)_q=\frac{\left(1-\alpha-\frac{\lambda \pi q^2}{S}\right)}{4\sqrt{\pi\,S (1+\ell)}}=\frac{1}{4\pi r_h}\frac{\left(1-\alpha-\frac{\lambda q^2}{r_h^2}
\right)}{\sqrt{1+\ell}},\nonumber\\
    \Phi&=\left(\frac{\partial \mathcal{M}}{\partial q}\right)_{S}=\frac{\lambda}{\sqrt{1+\ell}}\,\sqrt{\frac{\pi}{S}}\,q=\frac{\lambda}{\sqrt{1+\ell}}\,\frac{q}{r_h}.\label{bb9}
\end{align}

It can be readily shown that the above thermodynamic quantities fully satisfy both the differential form of the first law of black hole thermodynamics given in (\ref{bb8a}) and the Bekenstein-Smarr relation given by
\begin{equation}
    \mathcal{M}=2 T S+\Phi\,q.\label{bb10}
\end{equation}

The Helmholtz free energy is obtained as,
\begin{equation}
    \mathcal{F}=\mathcal{M}-T S=\frac{r_h}{4\sqrt{1+\ell}}
\left[1-\alpha+3\lambda \frac{q^2}{r^2_h}
\right].\label{bb11}
\end{equation}

Next, we determine the specific heat capacity for the thermodynamic system. It is defined using Eq. (\ref{bb8a}) by
\begin{equation}
    C_q=\left(\frac{d\mathcal{M}}{dT}\right)_{q}=T\left(\frac{dS}{dT}\right)_{q}.\label{bb12}
\end{equation}
Substituting $T$ form Eq. (\ref{bb9}) and $S=\pi r_h^2$, we obtain the following expression:
\begin{equation}
C_q=-
\frac{2\pi r_h^2\left(1-\alpha-\lambda\frac{q^2}{r_h^2}\right)}
{\left(1-\alpha-3 \lambda\frac{q^2}{r_h^2}\right)}.\label{bb13}
\end{equation}

Finally, the Gibbs free-energy is obtained as,
\begin{equation}
G=\mathcal{M}-T\,S-q \Phi=\frac{r_h}{4\sqrt{1+\ell}}
\left[1-\alpha-\lambda\frac{q^2}{r_h}
\right].\label{bb14}
\end{equation}

From the above thermodynamic analysis, we observed that the Lorentz-violating parameter $\ell$ and the string cloud parameter $\alpha$ jointly influenced the thermodynamic quantities, thereby modifying the results in comparison with the standard charged black hole case.\\

In Fig.~\ref{fig1Helmholtz}, the Helmholtz free energy  plotted against the horizon radius  for various values of $\ell$ and $\alpha$. The curves indicate that the free energy first decreases, reaches a minimum at small $r_{h}$, and then increases steadily for larger $r_{h}$, suggesting the presence of a thermodynamically preferred configuration near the minimum. Variations in the LV parameter $\ell$ and the CoS parameter $\alpha$ cause the curves to shift downward, indicating that an increase in either $\ell$ or $\alpha$ lowers the Helmholtz free energy. Fig.~\ref{fig1SpecificHeat} shows the specific heat $C_{q}$ as a function of $r_{h}$ where a divergence occurs at a critical radius. This divergence marks the division between the regions of negative and positive specific heat, which correspond to thermodynamically unstable small black holes and stable large black holes, respectively, and indicates a second-order phase transition. The position of the divergence changes with the parameters $\ell$ and $\alpha$, showing their influence on the stability structure of the system. Fig.~\ref{fig1Gibbs} presents the variation of the Gibbs free energy G with respect to he horizon radius  for various values of $\ell$ and $\alpha$. It is observed that the Gibbs free energy increases monotonically with increasing $r_{h}$, indicating that larger black holes correspond to higher thermodynamic potential within the considered parameter regime. In the left panel, with $\alpha$ held constant, increasing the LV parameter $\ell$ results in a small decrease in the Gibbs free energy for a given horizon radius. Similarly, the right panel shows that increasing the CoS parameter $\alpha$ also lowers the Gibbs free energy when $\ell$ is fixed. This behavior suggests that both the LV and CoS parameters suppress the thermodynamic potential of the system, which modifies the thermodynamic characteristics of the black hole.\\
In the limit $\ell=0$, corresponding to the absence of Lorentz-violating effects and $\alpha=0$, corresponding to the absence of string cloud, the selected space-time reduces to RN black hole solution. In this limit, the thermodynamic results simplify as
\begin{align}
    r_h&=M \left(1+\sqrt{1-q^2/M^2}\right),\nonumber\\
    T&=\frac{1}{4\pi r_h}\left(\frac{M}{r_h}-\frac{q^2}{r_h^2}\right),\nonumber\\
    \Phi&=q/r_h,\nonumber\\
    F&=\frac{r_h}{4}\left(1+3\frac{q^2}{r_h^2}\right),\nonumber\\
    C_q&=-2 \pi r_h^2\,\frac{r_h^2-q^2}{r_h^2-3 q^2},\nonumber\\
    G&=\frac{r_h}{4}\left(1-\frac{q^2}{r_h^2}\right)\label{reduce}
\end{align}
which are similar to the thermodynamic results for the RN black hole case.

Fig.~\ref{ThermodynamicsComprision} shows a comparative analysis of the thermodynamic properties of the charged black hole in bumblebee gravity with a surrounding CoS with other well-known cases to which it reduces in specific limits. Panel (a) presents the Hawking temperature as a function of the horizon radius. The Schwarzschild black hole follows the expected inverse relationship where temperature $T$ is proportional to $1/r_{h}$. In contrast, the RN black hole displays a non-monotonic temperature pattern because of its electric charge. The Letelier black hole changes the temperature profile by including the influence of the string cloud parameter. The bumblebee black hole with CoS shows a lower peak temperature that happens at a slightly different horizon radius. This indicates that the LV parameter reduces the Hawking temperature compared to the RN and Schwarzschild black holes. Panel (b) compares the Helmholtz free energy for the different black hole models. All cases show similar qualitative behavior, but the presence of the LV parameter shifts the free-energy curve downward compared to the RN and Schwarzschild black holes. This shift suggests that Lorentz-violating effects reduce the thermodynamic potential and alter the energetically favored configurations. Panel (c) shows the specific heat $C_{q}$. The Schwarzschild black hole has negative specific heat and is thermodynamically unstable. In contrast, the RN and Letelier black holes have a divergence in $C_{q}$, indicating a second-order phase transition between small and large black hole phases. In the bumblebee gravity case, the divergence point shifts, showing that the LV parameter changes where the phase transition occurs and affects the stability range of the system. Finally, panel (d) presents the Gibbs free energy $G$ as a function of $r_h$.  The overall monotonic trend remains consistent across the models, but the bumblebee black hole with CoS shows consistently lower values compared to the RN and Schwarzschild cases. The Lorentz-violating bumblebee field introduces quantitative deviations in the thermodynamic potentials, changing the overall thermodynamic behavior of the black hole spacetime.

\section{Optical Properties}\label{sec:4}

The optical properties of a black hole can be explored by studying the motion of photons along null geodesics. Using the Lagrangian formalism, one can derive a one-dimensional equation of motion for equatorial trajectories and obtain the corresponding effective potential. This effective potential provides crucial information about the photon sphere, photon trajectories, and the forces experienced by photons in the given gravitational field~\cite{chandrasekhar1984,Wald1984}. Recently, the Event Horizon Telescope (EHT) achieved the first direct imaging of the supermassive black holes M87* at the center of the Virgo A galaxy and Sgr~A* at the center of the Milky Way. These groundbreaking observations make the study of black hole shadows an especially timely and important endeavor.

We consider the null geodesic motion in the equatorial plane defined by $\theta=\pi/2$ and $\dot{\theta}=0$ (see for example \cite{Al-Badawi2026,ALBADAWI2025185,Kukreti:2025rzn,Kala:2022uog}). Using the condition $ds=0$ for the metric (\ref{bb1}) yields
\begin{equation}
    -f(r)\,\dot{t}^2+\frac{1+\ell}{f(r)}\,\dot{r}^2+r^2\,\dot{\phi}^2=0,\label{dd1}
\end{equation}

The space-time is independent on $t,\,\phi$ thus there exist two conserved quantities known as the conserved energy ($\mathrm{E}$) and the conserved angular momentum ($\mathrm{L}$). These are given by
\begin{equation}
    \dot{t}=\frac{\mathrm{E}}{f(r)},\qquad \dot{\phi}=\frac{\mathrm{L}}{r^2}.\label{dd2}
\end{equation}

With these, the geodesics equation for the radial coordinate $r$ from Eq. (\ref{dd1}) becomes
\begin{equation}
    \dot{r}^2=V_{\rm eff},\label{dd3}
\end{equation}
where the effective potential is given by
\begin{equation}
    V_\text{eff}(r)=\frac{1}{1+\ell}\left(\mathrm{E}^2-\frac{\mathrm{L}^2}{r^2}\,f(r)\right).\label{dd4}
\end{equation}

Now, we study the circular orbits motion and discuss the relevant quantities associated with these. For circular orbits of radius $r=r_c$, we have the conditions $\dot{r}=0$ and $\ddot{r}=0$. Thereby, using (\ref{dd3}), we find the following two relations
\begin{equation}
    \mathrm{E}^2=\frac{\mathrm{L}^2}{r^2}\,f(r),\quad\quad V'_\text{eff}(r)=0.\label{dd5}
\end{equation}

The first relation $V_\text{eff}(r)=\mathrm{E}^2$ gives us the critical impact parameter for photon particle and is given by
\begin{equation}
    \frac{1}{\beta_c}=\frac{\mathrm{E}}{\mathrm{L}}=\frac{\sqrt{1-\alpha-\lambda\,\frac{q^2}{r^2}}}{r}\Bigg{|_{r=r_c}}.\label{dd6}
\end{equation}

It is clear from Eq. (\ref{dd6}) that the impact parameter of photons arriving from infinity and reaching a closest approach before being deflected by the black hole depends on multiple parameters. Both the string cloud parameter $\alpha$, LSB parameter $\ell$, the electric charge $q$, and the black hole mass $M$.

The second relation $V'_\text{eff}(r)=0$ gives us the photon sphere radius $r=r_\text{s}$ given by the following equation
\begin{equation}
    (1-\alpha)\,r_s^2-3M r_s+2\lambda\,q^2=0.\label{dd7}
\end{equation}
The exact  expression for the photon sphere radius is
\begin{equation}
   r_\text{s}=\frac{3 M+\sqrt{9 M^2-8 \lambda\,(1-\alpha)\,q^2}}{2(1-\alpha)}.\label{dd8} 
\end{equation}
The photon sphere exist provided we have the following constraint  
\begin{equation}
  q^2 <\frac{9 M^2}{8\,\lambda\,(1-\alpha)}.\label{constraint}  
\end{equation}

\begin{itemize}
    \item When $\ell=0$ and $\alpha=0$, the photon sphere radius simplifies as,
    \begin{equation}
   r_\text{s}=\frac{3 M+\sqrt{9 M^2-8 q^2}}{2}\label{dd8a} 
\end{equation}
which is analogue to the Reissner-Nordstrom black hole solution \cite{Eiroa2002}.

\item When $q=0$, the photon sphere radius simplifies as,
\begin{equation}
   r_\text{s}=\frac{3 M}{1-\alpha}\label{dd8b} 
\end{equation}
which is similar to the Letelier black hole case.

\item When $\alpha=0$, the photon sphere radius simplifies as,
\begin{equation}
   r_\text{s}=\frac{3 M+\sqrt{9 M^2-8 \lambda q^2}}{2}.\label{dd8c} 
\end{equation}
\end{itemize}

\begin{figure*}[ht!]
	\centerline{
	\includegraphics[width=120mm,height=80mm]{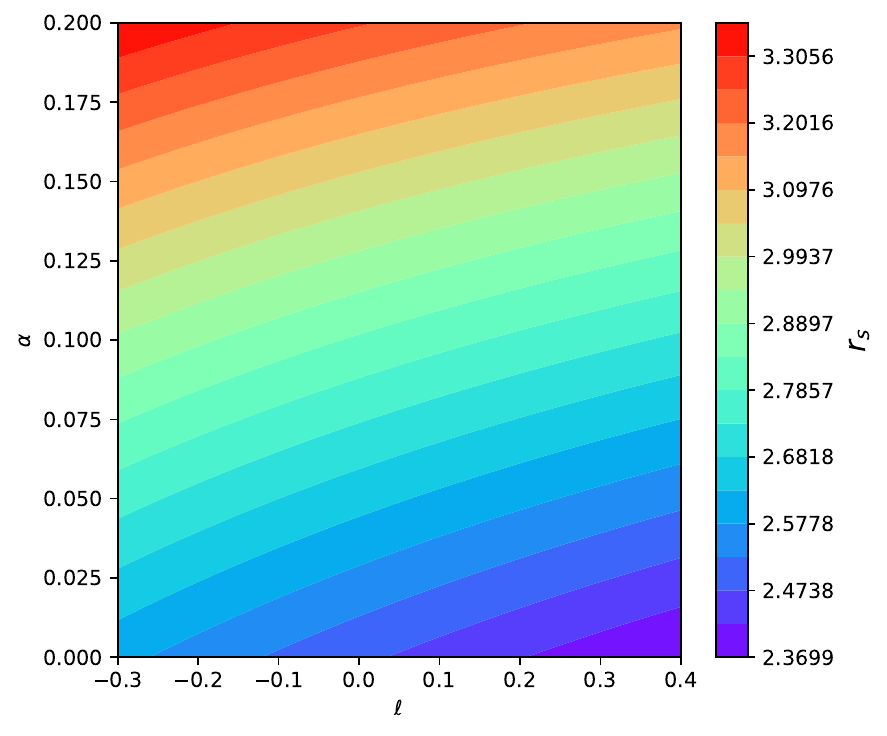}
    }
	\caption{Two-dimensional contour plot showing the variation of black hole photon sphere with $\ell$ and $\alpha$. Here we fix, $M=1$ and $q=0.8$.}
\label{figPhotonSphere2D}
\end{figure*}

Fig.~\ref{figPhotonSphere2D} illustrates a two--dimensional contour plot of the photon sphere radius $r_s$ as a function of the LV parameter $\ell$ and the cosmological parameter $\alpha$, while keeping the mass and charge fixed at $M=1$ and $q=0.8$, respectively. The color bar represents the magnitude of the photon sphere radius, with warmer colors (red and orange) corresponding to larger values of $r_s$ and cooler colors (blue and purple) indicating smaller values. It is evident from the figure that the photon sphere radius increases monotonically with increasing $\alpha$, whereas it decreases as the bumblebee parameter $\ell$ increases. This behavior demonstrates that the cosmological correction tends to expand the photon sphere, while the Lorentz symmetry breaking effects encoded in $\ell$ lead to a contraction of the photon orbit.
The smooth and continuous contour lines indicate a regular dependence of the photon sphere radius on both parameters throughout the explored parameter space, with no abrupt transitions or discontinuities. This confirms the stability of the photon sphere structure within the chosen ranges of $\ell$ and $\alpha$. Overall, Fig.~5 highlights the combined influence of Lorentz-violating effects and cosmological corrections on the location of the photon sphere, emphasizing their significant role in modifying the optical properties and shadow characteristics of the black hole spacetime.

Next, we determine the shadow radius $R_{\rm sh}$ for a static observer located at position $r_O$. Following the method in \cite{Volker2022},  the
angular radius $\vartheta_{\rm sh}$ of the shadow is give by
\begin{equation}
    \sin \vartheta_{\rm sh}=\frac{h(r_s)}{h(r_O)},\qquad h(r)=\frac{r}{\sqrt{f(r)}}.\label{dd9}
\end{equation}
In our case at hand, we find this angular radius as
\begin{equation}
    \vartheta_{\rm sh} =\frac{r_s}{r_O}\,\sqrt{\frac{1-\alpha-\frac{2M}{r_O}+\lambda\,\frac{q^2}{r^2_O}}{1-\alpha-\frac{2M}{r_s}+\lambda\,\frac{q^2}{r^2_s}}}.\label{dd10}
\end{equation}
Therefore, the shadow radius is given by
\begin{equation}
    R_{\rm sh} \simeq \vartheta_{\rm sh}\,r_O=r_s\,\sqrt{\frac{1-\alpha-\frac{2M}{r_O}+\lambda\,\frac{q^2}{r^2_O}}{1-\alpha-\frac{2M}{r_s}+\lambda\,\frac{q^2}{r^2_s}}}.\label{dd11}
\end{equation}

\begin{figure*}[ht!]
	\centerline{
		\includegraphics[width=80mm,height=70mm]{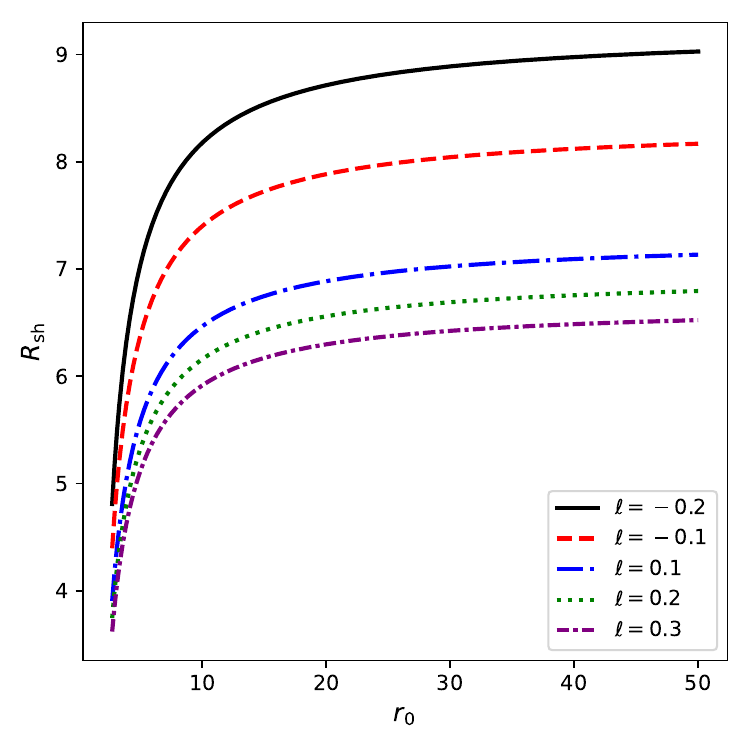}\qquad
        \includegraphics[width=80mm,height=70mm]{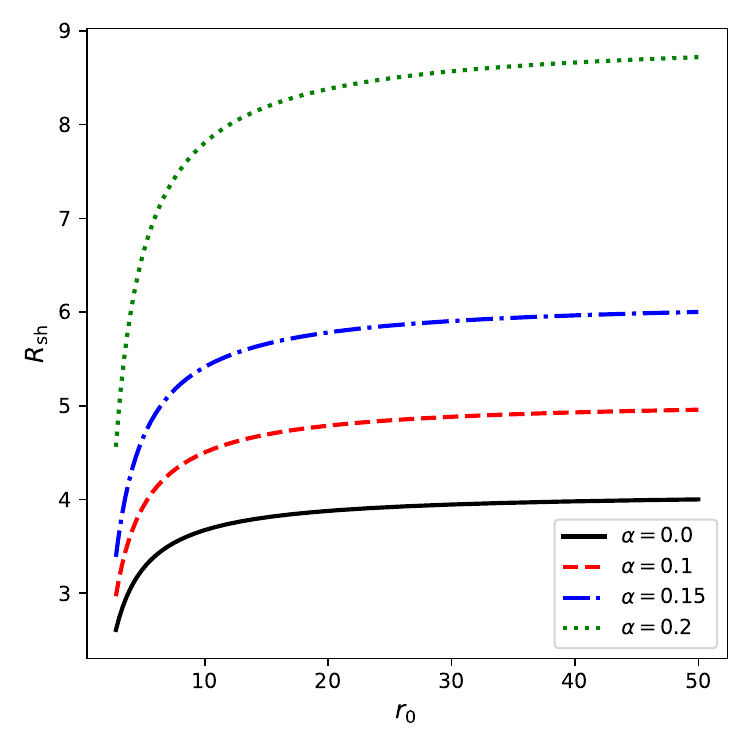}}(a) $\alpha=0.1$ \hspace{5cm} (b) $\ell=0.2$
	\caption{The variation of angular radius of black hole shadow as a function of $r_{0}$; for different values of LV parameter ($\ell$) and CoS parameter ($\alpha$). Here we fix, $M=1$, $q=0.8$, $\ell=0.2$ and $\alpha=0.1$ where applicable.}
\label{fig1}
\end{figure*}

For a static distant observer, the shadow radius simplifies as,
\begin{equation}
    R_{\rm sh} = r_s\,\sqrt{\frac{1-\alpha}{1-\alpha-\frac{2M}{r_s}+\lambda\,\frac{q^2}{r^2_s}}}.\label{dd12}
\end{equation}

\begin{figure*}[ht!]
	\centerline{
		\includegraphics[width=80mm,height=90mm]{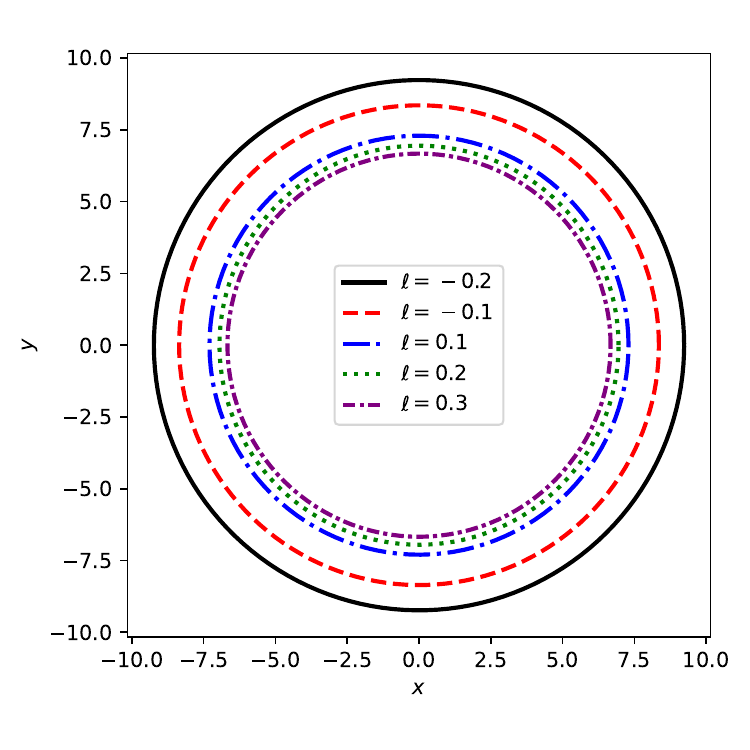}
        \includegraphics[width=80mm,height=90mm]{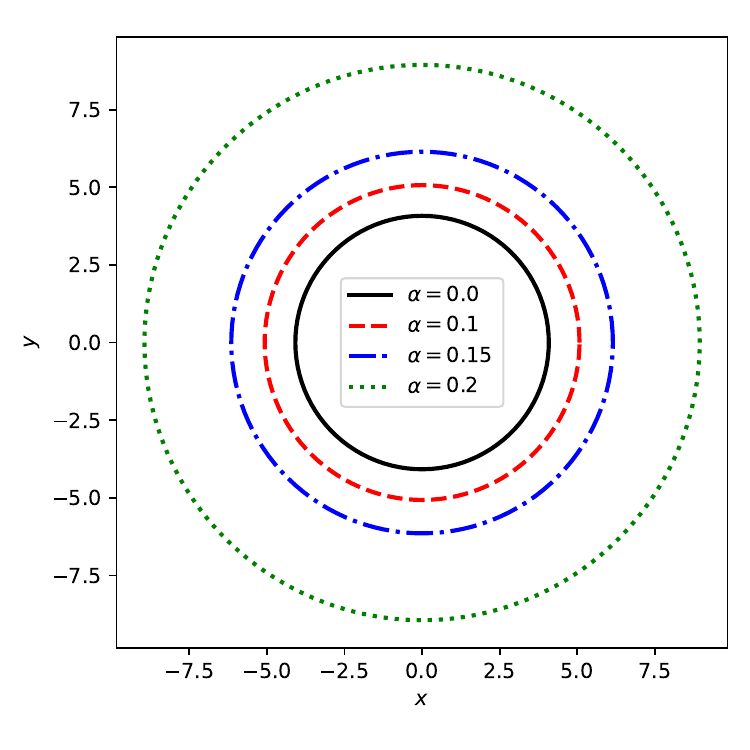}}(a) $\alpha=0.1$ \hspace{5cm} (b) $\ell=0.2$
	\caption{The parametric plot of angular radius of black hole shadow in $(\mathrm x, \mathrm y)$ plane; for different values of LV parameter ($\ell$) and CoS parameter ($\alpha$). Here we fix, $M=1$, $q=0.8$, $\ell=0.2$ and $\alpha=0.1$ where applicable.}
\label{fig2}
\end{figure*}

A few special cases are as follows:

\begin{itemize}
    \item $\ell=0$ and $\alpha=0$,\qquad \(R_{\rm sh} =\frac{3\sqrt{3} M}{2\sqrt{2}}\,\frac{\left(1+\sqrt{1-\frac{8q^2}{9M^2}}\right)^2}{\sqrt{1-\frac{2 q^2}{3M^2}+\sqrt{1-\frac{8q^2}{9M^2}}}}.\)

    \item $q=0$,\qquad \(R_{\rm sh} =\frac{3\sqrt{3} M}{(1-\alpha)^{3/2}}.\)

    \item $\alpha=0$,\qquad \( R_{\rm sh} = \frac{3\sqrt{3} M}{2\sqrt{2}}\,\frac{\left(1+\sqrt{1-\lambda\frac{8q^2}{9M^2}}\right)^2}{\sqrt{1-\lambda\frac{2 q^2}{3M^2}+\sqrt{1-\lambda\frac{8q^2}{9M^2}}}}.\)

\end{itemize}
In Fig.~\ref{fig1}, the angular radius of the black hole shadow is plotted against $r_{0}$ for various values of the parameters $\ell$ and $\alpha$. The analysis shows that when the static observer is at a finite distance from the black hole, the shadow radius reaches its maximum value. As the observer moves farther away, the shadow radius gradually saturates and approaches a constant asymptotic value. It is observed that an increase in the LV parameter $\ell$ leads to a reduction in the shadow size. In contrast, the analysis for the CoS parameter alpha indicates that the shadow radius increases with increasing alpha. These results show that the two parameters have opposite effects on the black hole shadow radius. This study shows that the bumblebee field and CoS parameters have competing effects on the size of the black hole shadow. This finding suggests a way to use high-resolution VLBI observations to constrain these parameters.\\
The parametric plot of angular radius of black hole shadow in $(\mathrm x, \mathrm y)$ plane; for different values of LV parameter ($\ell$) and CoS parameter ($\alpha$) is depicted in Fig.~\ref{fig2}. In left panel, we plot the shadow radius for different values of $\ell$. It has been observed that the the angular size of black hole shadow attains its maximum at lowest values of $\ell$. The shadow size reduces with an increase in LV parameter. The obtained results suggests that the bumblebee parameter effectively weakens the light bending strength by modifying the spacetime geometry, leading to a smaller photon capture region and hence a reduced shadow radius. On the other hand at right panel we plot shadow radius for different values of CoS parameter. The shadow radius increases with an increase in CoS parameter, the shadow radius approach minimum for $\alpha=0$. Physically, a larger CoS parameter enhances the effective gravitational potential experienced by photons, thereby enlarging the photon sphere and producing a larger shadow.\\

We now constrain the parameters $\alpha$ and $\ell$ using the EHT observational results for Sgr A*. The numerical bounds adopted from these observations are summarized in Table~\ref{shadow_bounds}.
\begin{table}[h]
\centering
\caption{Constraints on the shadow radius of Sgr A* obtained from EHT observations using Keck and VLTI mass-to-distance measurements~\cite{EHTL12}.}
\begin{tabular}{c c}
\hline
\textbf{Confidence Level} & $\mathbf{R_{\rm sh}/M}$ \\
\hline
$1\sigma$ bound & $4.55 \le R_{\rm sh}/M \le 5.22$ \\
$2\sigma$ bound & $4.21 \le R_{\rm sh}/M \le 5.56$ \\
\hline
\end{tabular}
\label{shadow_bounds}
\end{table}

\begin{figure*}[tbhp]
	\centerline{
		\includegraphics[width=180mm,height=140mm]{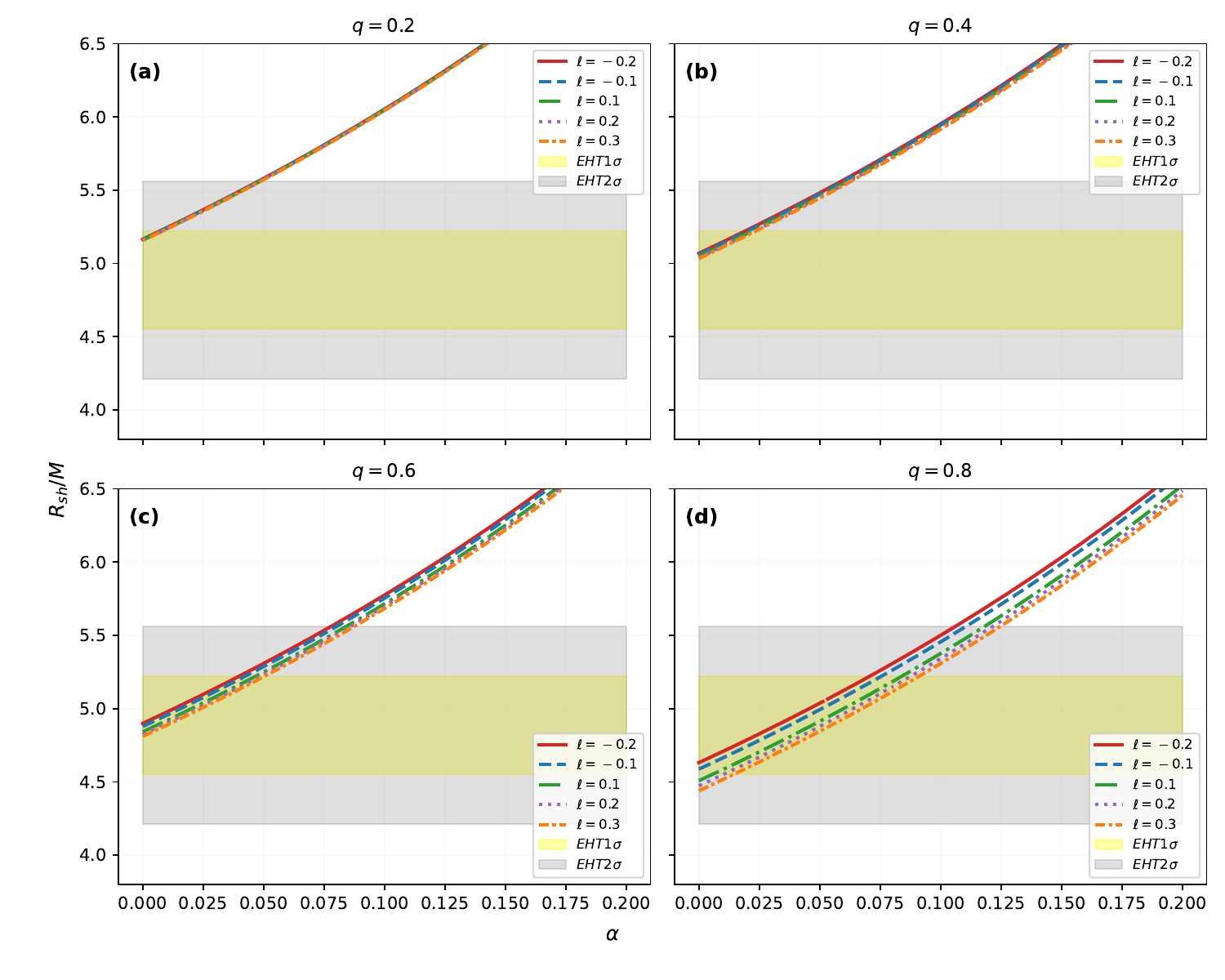}}
	\caption{Variation of the shadow radius $R_{\rm sh}/M$ as a function of the parameter $\alpha$ for different values of the charge parameter $q$. The colored curves represent different values of the LV parameter $\ell$. The shaded yellow and gray regions denote the $1\sigma$ and $2\sigma$ confidence intervals obtained from the EHT VLBI observations of Sgr~A*, respectively.}
\label{fig3}
\end{figure*}

Fig.~\ref{fig3} shows how the black hole shadow radius $R_{\rm sh}/M$ depends on the parameter $\alpha$ for various fixed values of the charge parameter $q$. Each panel represents a different value of $q$: specifically, $q=0.2$ (a), $q=0.4$ (b), $q=0.6$ (c), and $q=0.8$ (d). The colored curves indicate different choices of the LV parameter $\ell$. The shaded yellow and gray areas represent the $1 \sigma$ and $2 \sigma$ confidence levels based on the Event Horizon Telescope VLBI observational constraints on the shadow size of Sgr A*. These observational bands set strict limits on the allowed parameter space of the model. From the figure, it is clear that the shadow radius increases continuously with the parameter $\alpha$ for all values of $q$ and $\ell$ considered. For smaller values of $q$, the curves fall well within the $1 \sigma$ and $2 \sigma$ observational bounds, showing good alignment with the EHT data. As $q$ increases, the deviations from the observational bands become more noticeable, especially for larger values of $\alpha$, which imposes stronger limits on the acceptable range of the model parameters. For specific parameter choices, the resulting constraints are summarized in Table~\ref{shadow_constarints}. In summary, comparing the EHT VLBI observations of Sgr A* helps us limit the parameters $\alpha$, $q$, and $\ell$, revealing that only a small area of the parameter space remains consistent with current observational data.

\begin{table}
\centering
\caption{Constraints on the $\rm CS$ parameter $\alpha$ for different values of the $\rm LV$ parameter $\ell$ using Sgr A* observations data. Here, we fix $q=0.8$.}
\begin{tabular}{|c|c|c|c|}
\hline
Case & Parameter & $1\sigma$ upper bound & $2\sigma$ upper bound \\ 
\hline
\multirow{2}{*}{Sgr A*} 
& $\ell=-0.2$ & 0.063 & 0.106 \\ 
& $\ell=0.3$  & 0.087 & 0.125 \\ 
\hline
\end{tabular}
\label{shadow_constarints}
\end{table}

\section{Impact of In-falling Accretion Gas on Shadow}\label{sec:5}

In this subsection, we investigate the influence of a radially infalling accretion flow on black hole shadow images within a generalized spacetime framework, focusing on the role of the model parameters. To describe the appearance of a black hole surrounded by accreting matter, we assume an optically thin, radially infalling gas distribution and compute the radiation detected by a distant observer located on the image plane $(\mathrm x, \mathrm y)$.

\noindent The specific intensity observed at frequency $\nu_{\text{obs}}$ is obtained by integrating the emissivity along the photon trajectory $\gamma$~\cite{Jaroszynski:1997bw},
\begin{equation}
I_{\text{obs}}(\nu_{\text{obs}}, \mathrm x, \mathrm y)
= \int_{\gamma} g^{3}\, j_{\nu_{\text{em}}}\, d\ell_{\text{prop}},
\end{equation}
where $g=\nu_{\text{obs}}/\nu_{\text{em}}$ is the redshift factor, $j_{\nu_{\text{em}}}$ denotes the emissivity measured in the rest frame of the emitter, and $d\ell_{\text{prop}}$ is the proper length element along the photon path.

For a photon with four-momentum $p^{\mu}$, the redshift factor can be written as
\begin{equation}
g = \frac{p_{\mu} u_{\text{obs}}^{\mu}}{p_{\nu} u_{\text{em}}^{\nu}}
= \frac{-p_{t}}{-p_{t}u^{t}_{\text{em}} + p_{r}u^{r}_{\text{em}}},
\end{equation}
where the four-velocity of a static observer at infinity is given by
$u_{\text{obs}}^{\mu}=(1,0,0,0)$.

\noindent We assume that the accretion flow undergoes radial free fall in the equatorial plane $(\theta=\pi/2)$. The corresponding four-velocity components of the emitter are
\begin{align}
u^{t}_{\text{em}} &= \frac{E}{|g_{tt}(r)|}, \\
u^{r}_{\text{em}} &= -\sqrt{E^{2} - |g_{tt}(r)|},
\end{align}
where the metric function $g_{tt}(r)$ depends on the LV parameter $\ell$, the CoS parameter $\alpha$ and charge parameter $q$ through the lapse function of the spacetime.

\noindent For photons propagating in the equatorial plane with impact parameter $b=L/E$, the momentum components are
\begin{align}
p_{t} &= -E, \\
p_{\phi} &= bE, \\
p_{r} &= \pm E\sqrt{1 - \frac{b^{2} f(r)}{r^{2}}},
\end{align}
where the radial turning point determines the photon sphere and the associated critical impact parameter $b_{\text{ph}}$.

\noindent The emissivity of the accreting flow is assumed to be monochromatic and decreases with radius according to~\cite{Gan:2021pwu}
\begin{equation}
j_{\nu_{\text{em}}}(r)
= j_{0}\,\delta(\nu_{\text{em}} - \nu_{0})\, r^{-2},
\end{equation}
which is appropriate for an optically thin, radially infalling medium.

\noindent The proper length element along the photon trajectory is expressed as
\begin{equation}
d\ell_{\text{prop}}
= \frac{p_{t}}{p_{r}} \sqrt{g_{rr}(r)}\, dr
= \frac{E}{|p_{r}|} \sqrt{g_{rr}(r)}\, dr.
\end{equation}
After integrating over all emitted frequencies, the total observed intensity on the image plane becomes~\cite{Kala:2025fld},
\begin{equation}
I_{\text{obs}}(\mathrm x, \mathrm y)
\propto \int_{\gamma}
\frac{g^{3} p_{t}}{r^{2} |p_{r}|}\,
\sqrt{g_{rr}(r)}\,
dr .
\end{equation}
This expression describes the brightness distribution of the black hole image produced by an optically thin, radially infalling accretion flow and allows us to examine how the LV parameter $\ell$ and the CoS parameter $\alpha$ influence the observed intensity and shadow morphology.

\begin{figure*}[tbhp]
	\centerline{
		\includegraphics[width=180mm,height=160mm]{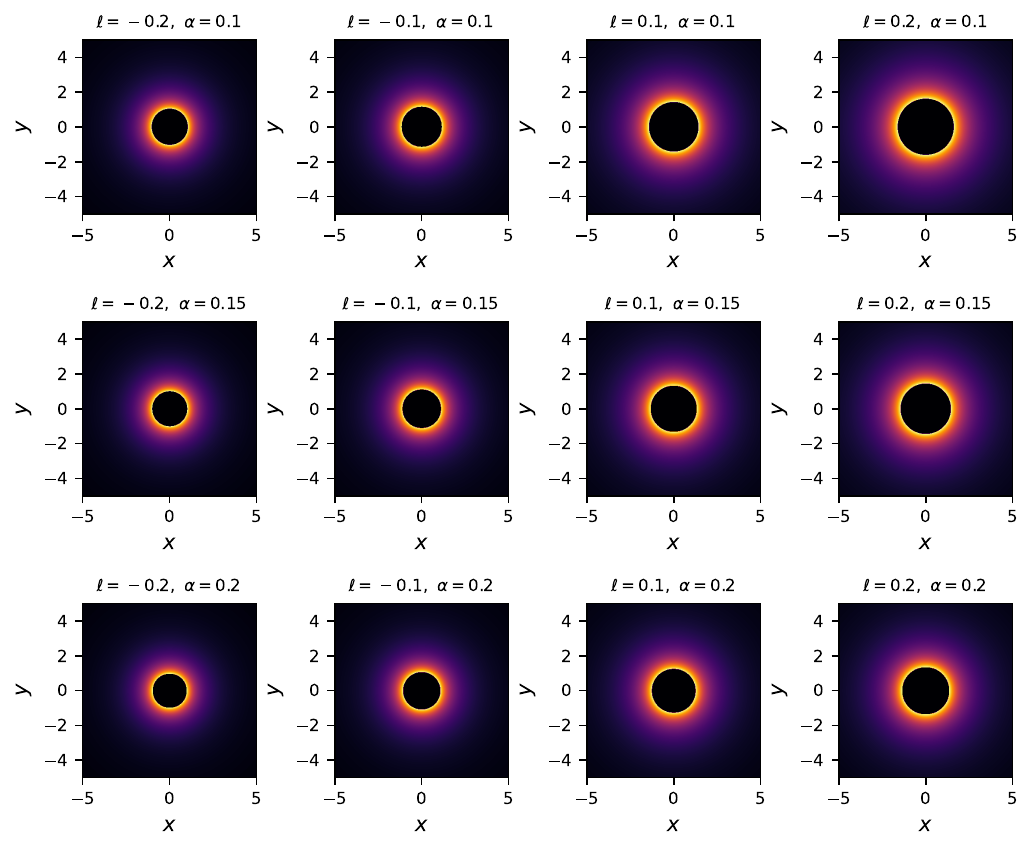}}
	\caption{Images of the optically thin emission region surrounding the black hole for different values of the parameters $\ell$ and $\alpha$. The $\mathrm x$ and $\mathrm y$ axes represent the angular celestial coordinates in the observer’s sky. Here, we fix $q=0.8$.}
\label{figShadowImageAccretion}
\end{figure*}

Fig.~\ref{figShadowImageAccretion} illustrates the optically thin emission images of the black hole shadow for different combinations of the LV parameter $\ell$ and the CoS parameter $\alpha$. The panels are arranged such that $\ell$ varies horizontally for fixed values of $\alpha$, while $\alpha$ increases from top to bottom. The dark central region corresponds to the black hole shadow, which is bounded by the photon sphere and surrounded by a bright emission ring produced by photons orbiting near unstable circular trajectories. It is observed that variations in $\ell$ significantly modify the apparent size of the shadow and the thickness of the luminous ring. For smaller values of $\ell$, the shadow appears larger, whereas increasing $\ell$ leads to a gradual reduction in the shadow radius, indicating that Lorentz symmetry breaking encoded in the LV parameter affects the null geodesic structure and the location of the photon sphere. On the other hand, increasing the CoS parameter $\alpha$ leads to a systematic decrease in the shadow size and a suppression of the surrounding emission intensity. This behavior reflects the influence of the surrounding CoS medium on photon propagation, which effectively modifies the gravitational potential and reduces the critical impact parameter for photon capture. In conclusion, the figure demonstrates that both parameters $\ell$ and $\alpha$ leave distinct imprints on the morphology of the black hole shadow and the brightness distribution of the optically thin emission region. These results confirm that the shadow boundary is primarily governed by the spacetime geometry, while the observed intensity profile is sensitive to the surrounding medium. Consequently, precise observations of black hole shadows may provide a potential probe for testing Lorentz-violating effects and environmental influences described by the LV and CoS parameters.

\section{Solar System Tests}\label{sec:6}

General Relativity  has undergone extensive testing, particularly in the weak-field regime, such as within the Solar System. So far, no observations have contradicted the theory. Nevertheless, it is useful to evaluate the potential effects of the present Lorentz-violating theory by comparing its derived solutions with Solar System observations. The theoretical parameter $\ell$ can be constrained by matching the theory's predictions to experimental results. Since the influence of the cosmological constant is negligible on Solar System scales, we focus here solely on the RN-like metric~(\ref{metric}).

The motion of a test particle of mass $m$ along its geodesics can be described by the Lagrangian
\begin{equation}
\mathbb{L} = \frac{1}{2} m g_{\mu\nu} \frac{dx^{\mu}}{d\tau} \frac{dx^{\nu}}{d\tau}, \label{ss1}
\end{equation}
where the dot represents the derivative with respect to the affine parameter $\tau$. From the normalization conditions for the four-velocity of timelike null particles, one can express the Lagrangian density as \(\mathbb{L} = \frac{\epsilon}{2}.\) Accordingly, for massive particles, $\epsilon =-1$.

Since the spacetime is static and spherically symmetric, the Lagrangian is independent of $t$ and $\phi$. As a result, there are two conserved quantities: the energy $\mathcal{E}$ and the angular momentum $\mathcal{L}$ per unit mass of test particles, given by
\begin{align}
\mathcal{E}=f(r)\, \frac{dt}{d\tau},\qquad
\mathcal{L}= r^2\, \frac{d\phi}{d\tau}.\label{ss2}
\end{align}

Then, from the conserved quantities in Eq.~(\ref{ss2}), for timelike geodesics, one can obtain a single differential equation for the radial coordinate $r$ in terms of the proper time $\tau$:
\begin{equation}
(1+\ell)\,\left(\frac{dr}{d\tau}\right)^2+\left(1+\frac{\mathcal{L}^2}{r^2}\right)\,f(r)= \mathcal{E}^2.\label{ss3}
\end{equation}

We now introduce the variable $v(\phi)=\frac{1}{r(\phi)}$, so that
\begin{equation}
\frac{dr}{d\tau}= \frac{dr}{d\phi}\frac{d\phi}{d\tau}= - \mathcal{L}\,\frac{dv}{d\phi}. \label{ss4}
\end{equation}
By substituting this into Eq.~(\ref{ss3}), we obtain
\begin{align}
(1+\ell) \left(\frac{dv}{d\phi}\right)^2 + \left(1-\alpha+\lambda\frac{q^2}{\mathcal{L}^2}\right) v^2 &= \frac{\mathcal{E}^2 - 1+\alpha}{\mathcal{L}^2} + \frac{2 M}{\mathcal{L}^2} v \nonumber\\
&+ 2 M v^3-\lambda q^2 v^4. \label{ss5}
\end{align}

As is usually done in this treatment, it is preferable to solve the second-order differential equation obtained by differentiating the above equation with respect to $\phi$:
\begin{equation}
(1+\ell) \frac{d^2v}{d\phi^2}+\left(1-\alpha+\lambda\frac{q^2}{\mathcal{L}^2}\right) v= \frac{M}{\mathcal{L}^2}+ 3 M v^2-2\lambda q^2 v^3. \label{ss6}
\end{equation}

In order to solve Eq.~(\ref{ss6}), we set the electric charge $q=0$ for simplicity. In this limiting case, Eq.~(\ref{ss6}) reduces to
\begin{equation}
(1+\ell) \frac{d^2v}{d\phi^2} + (1-\alpha) v = \frac{M}{\mathcal{L}^2} + 3 M v^2. \label{ss7}
\end{equation}

Here, the Lorentz-violating (LV) parameter $\ell$ appears in the coefficient of the first term, while the string cloud contribution modifies the coefficient of the second term, preserving the overall structure familiar from GR (GR). To solve Eq.~(\ref{ss7}) perturbatively, and noting that $\ell \ll 1$, it is valid to treat the last term as a relativistic correction compared with the Newtonian case. The perturbative solution is expressed in terms of a small parameter $\epsilon = 3M^2/\mathcal{L}^2$:
\begin{equation}
v \simeq v^{(0)} + \epsilon v^{(1)}, \label{ss8}
\end{equation}
where $v^{(0)}$ satisfies the zeroth-order differential equation
\begin{equation}
(1+\ell) \frac{d^2v^{(0)}}{d\phi^2} + (1-\alpha) v^{(0)} - \frac{M}{\mathcal{L}^2} = 0. \label{ss9}
\end{equation}

The solution of Eq.~(\ref{ss9}) is
\begin{equation}
v^{(0)} = \frac{M}{\mathcal{L}^2} \left[ 1 + e \, \cos \left( \sqrt{\frac{1-\alpha}{1+\ell}} \, \phi \right) \right], \label{ss10}
\end{equation}
which is analogous to the Newtonian result. The integration constants are chosen as the orbital eccentricity $e$ (assumed small, similar to GR) and the initial condition $\phi_0 = 0$.

At first order in $\epsilon$, the differential equation reads
\begin{equation}
(1+\ell) \frac{d^2v^{(1)}}{d\phi^2} + (1-\alpha) v^{(1)} - \frac{\mathcal{L}^2}{M} (v^{(0)})^2 = 0, \label{ss11}
\end{equation}
which admits an approximate solution of the form
\begin{align}
v^{(1)} &\simeq \frac{M}{\mathcal{L}^2} \, e \, \phi \, \sqrt{\frac{1-\alpha}{1+\ell}} \, \sin \left( \phi \, \sqrt{\frac{1-\alpha}{1+\ell}} \right)\nonumber\\
&+ \frac{M}{\mathcal{L}^2} \left[ \frac{1+e^2}{2} - \frac{e^2}{6} \cos \left( 2 \phi \, \sqrt{\frac{1-\alpha}{1+\ell}} \right) \right]. \label{ss12}
\end{align}

For practical purposes, the second term in Eq.~(\ref{ss12}) can be neglected, as it consists of a constant shift and an oscillatory contribution that averages to zero.

Therefore, the perturbative solution (\ref{ss6}) reads
\begin{align}
    v &\simeq \frac{M}{\mathcal{L}^2}  \Bigg[1+ e \cos \left( \sqrt{\frac{1-\alpha}{1+\ell}} \phi \right)\nonumber\\
    &+\epsilon e \phi \sqrt{\frac{1-\alpha}{1+\ell}} \sin \left( \phi \sqrt{\frac{1-\alpha}{1+\ell}} \right) \Bigg].\label{ss13}
\end{align}

Because \(\epsilon \ll 1\), the perturbative solution (\ref{ss13}) can be rewritten in the form of an ellipse equation,
\begin{equation}
    v \simeq \frac{M}{\mathcal{L}^2} \left[1+ e \cos \left( \sqrt{\frac{1-\alpha}{1+\ell}} (1-\epsilon)\phi \right)\right].\label{ss14} 
\end{equation}

Despite the presence of Lorentz violation and a string cloud, the orbit remains periodic with a period $\Phi$,  
\begin{equation}
\Phi = 2 \pi \frac{\sqrt{1+\ell}}{\sqrt{1-\alpha} \, (1-\epsilon)} \approx 2 \pi + \Delta \Phi, \label{ss15}
\end{equation}
where $\Delta \Phi$ represents the advance of the perihelion.  

By expanding to first order in the small parameters $\ell$, $\alpha$, and $\epsilon$, the perihelion shift $\Delta \Phi$ is given by
\begin{equation}
\Delta \Phi = 2 \pi \epsilon + \pi \ell + \pi \alpha = \Delta \Phi_{\rm GR} + \delta \Phi_{\rm LV} + \delta \Phi_{\rm CoS}, \label{ss16}
\end{equation}
where each term corresponds to a distinct contribution:  

- The standard GR contribution:
\[
\Delta \Phi_{\rm GR} = 2 \pi \epsilon=\frac{6\pi G_N\,m}{c^2\,(1-e^2)\,a},
\]  
with $c$ being the speed of light, \(m\) the geometrical mass, \(e\) the orbital eccentricity and \(a\) being the semi-major axis of the orbital ellipse.

- The contribution due to spontaneous Lorentz-symmetry breaking:
\begin{equation}
\delta \Phi_{\rm LV} = \pi \ell, \label{ss17}
\end{equation}
- The contribution due to the presence of a string cloud:
\begin{equation}
\delta \Phi_{\rm CoS} = \pi \alpha. \label{ss18}
\end{equation}

Equation~(\ref{ss16}) clearly shows that Lorentz violation and the string cloud introduce additional corrections to the standard general relativistic perihelion advance.
 
\section{Bending of Light}\label{sec:7}

The bending of light by a gravitational field arises due to the curvature of spacetime around massive objects. Photons follow null geodesics, and their trajectories are deflected when passing near a black hole. The deflection angle depends on the black hole’s mass, spin, and any modifications to the geometry, such as those introduced by string clouds, Lorentz-violating effects in Bumblebee gravity etc. 

For photon particles, the trajectories correspond to null geodesics, $ds^{2}=0$. Restricting the motion to the equatorial plane $(\theta=\pi/2)$ and using the conserved quantities, namely the energy $E$ and the angular momentum $L$, we obtain the equation of motion for the radial coordinate $r$ as,
\begin{equation}
(1+\ell)\dot r^{2}+f(r)\frac{L^{2}}{r^{2}}=E^{2},
\label{radial2}
\end{equation}
where the dot denotes differentiation with respect to an affine parameter.

\noindent Introducing the variable $u=r^{-1}$ and writing $r\equiv r(\phi)$, Eq.~(\ref{radial2}) can be transformed into the following orbit equation
\begin{equation}
(1+\ell)\frac{d^{2}u}{d\phi^{2}}+(1-\alpha)u-3Mu^{2}+2\lambda q^{2}u^{3}=0.
\label{orbit2}
\end{equation}
In the limit $\ell\to0$ and $\alpha\to0$, Eq.~(\ref{orbit2}) reduces to the corresponding Reissner--Nordström result, while for $q\to0$ it reproduces the Schwarzschild-like solution in bumblebee gravity.

\noindent To obtain an approximate solution, we apply a perturbative method by assuming $Mu\ll1$ and $q^{2}u^{2}\ll1$. Thus, we write
\begin{equation}
u(\phi)\simeq u^{(0)}+3Mu^{(1)}+q^{2}u^{(2)}.
\label{perturb2}
\end{equation}

\noindent The zeroth-order equation reads
\begin{equation}
(1+\ell)\frac{d^{2}u^{(0)}}{d\phi^{2}}+(1-\alpha)u^{(0)}=0,
\label{zeroth2}
\end{equation}
whose solution is
\begin{equation}
u^{(0)}=\frac{1}{\beta}\sin\,\left(\frac{\sqrt{1-\alpha}}{\sqrt{1+\ell}}\phi\right),
\label{u0}
\end{equation}
where $\beta$ is an integration constant identified with the impact parameter. This result corresponds to a straight-line trajectory modified by the presence of the string cloud parameter $\alpha$ and the Lorentz symmetry breaking parameter $\ell$.

\noindent For simplicity we now consider $q=0$, solving the first-order perturbation equations for $u^{(1)}$, the general approximate solution for $u(\phi)$ can be written as
\begin{align}
u(\phi) \simeq &
\frac{1}{\beta}\sin\,\left(\frac{\sqrt{1-\alpha}}{\sqrt{1+\ell}}\phi\right)
+ \frac{M}{\beta^{2}}\Bigg[1+A\cos\,\left(\frac{\sqrt{1-\alpha}}{\sqrt{1+\ell}}\phi\right) \nonumber\\
& + \cos^{2}\,\left(\frac{\sqrt{1-\alpha}}{\sqrt{1+\ell}}\phi\right)\Bigg],
\label{solution2}
\end{align}
where $A$ is an arbitrary constant. In order to determine the deflection angle, we impose the boundary conditions: (i) the source is located at $r\rightarrow\infty$ such that $u\rightarrow0$ when $\phi=-\delta_{1}$, and (ii) the observer is also located at $r\rightarrow\infty$ such that $u\rightarrow0$ when $\phi=+\delta_{2}$. The total deflection angle is then given by $\delta=\delta_{1}+\delta_{2}$. 

\noindent  By applying these boundary conditions to Eq.~(\ref{solution2}) and considering the limits $\ell\ll1$, $\alpha\ll1$, and $\delta_{1},\delta_{2}\ll1$, we obtain
\begin{equation}
\delta_{1}=\frac{M}{\beta}(2+A),
\label{delta1}
\end{equation}
and
\begin{equation}
\delta_{2}=\frac{M}{\beta}(2-A)+\frac{\pi\ell}{2}+\frac{\pi\alpha}{2}.
\label{delta2}
\end{equation}

\noindent Hence, the total deflection angle is
\begin{equation}
\delta=\delta_{\rm GR}+\delta_{\rm CoS}+\delta_{\rm LV},
\label{totaldelta}
\end{equation}
where
\begin{align}
\delta_{\rm GR} &= \frac{4GM}{c^{2}\beta}, \label{grterm}\\
\delta_{\rm CoS} &= \frac{\pi\alpha}{2}, \label{alphaterm}\\
\delta_{\rm LV} &= \frac{\pi\ell}{2}. \label{lvterm}
\end{align}

\noindent In the appropriate limit, when $\ell\to0$ and $\alpha\to0$, our result reduces to the standard deflection angle predicted by GR

\noindent For a light ray grazing the Sun, we set $M=M_{\odot}$ and $\beta \simeq R_{\odot}$. In GR, the predicted deflection angle is $\delta_{\rm GR}=4GM_{\odot}/(c^{2}R_{\odot})\simeq1.7516687''$~\cite{beringer2012particle}. The current observational uncertainty in the measurement of light deflection is of the order $\sim 0.0001051''$~\cite{lambert2011improved}. Requiring that the corrections arising from Lorentz violation and the cloud of strings remain within this observational uncertainty, we impose the condition
\begin{equation}
\left|\delta_{\rm LV}+\delta_{\rm CoS}\right|<0.0001051''.
\label{bound}
\end{equation}

\noindent Using the previously reported constraint on the Lorentz symmetry breaking parameter,
$-1.1\times10^{-10}\le \ell \le 5.4\times10^{-10}$~\cite{Yang:2023wtu}, the above bound leads to the corresponding constraint on the cloud-of-strings parameter
\begin{equation}
-8.6\times10^{-10} \lesssim \alpha \lesssim 4.3\times10^{-10}.
\end{equation}
\begin{table}[h]
\centering
\caption{Constraints on the $\rm LV$ parameter and $\rm CoS$ parameter from Solar system~(light grazing by Sun) test.}
\begin{tabular}{c}
\hline
Solar Test: Light Deflection \\ 
\hline
Light bending prediction from GR: $1.7516687''$ \\ 
Current Observational uncertainty: $0.0001051''$ \\ 
Constraints on $\rm LV$ parameter: $-1.1\times10^{-10}\le \ell \le 5.4\times10^{-10}$ \\ 
Constraints on $\rm CoS$ parameter: $-8.6\times10^{-10}\lesssim \alpha \lesssim 4.3\times10^{-10}$ \\ 
\hline
\end{tabular}
\label{tab:solar_constraints}
\end{table}
\noindent These limits indicate that the string cloud parameter must also be extremely small in order to remain consistent with solar-system observations of light deflection.

\begin{figure*}[tbhp]
	\centerline{
		\includegraphics[width=180mm,height=140mm]{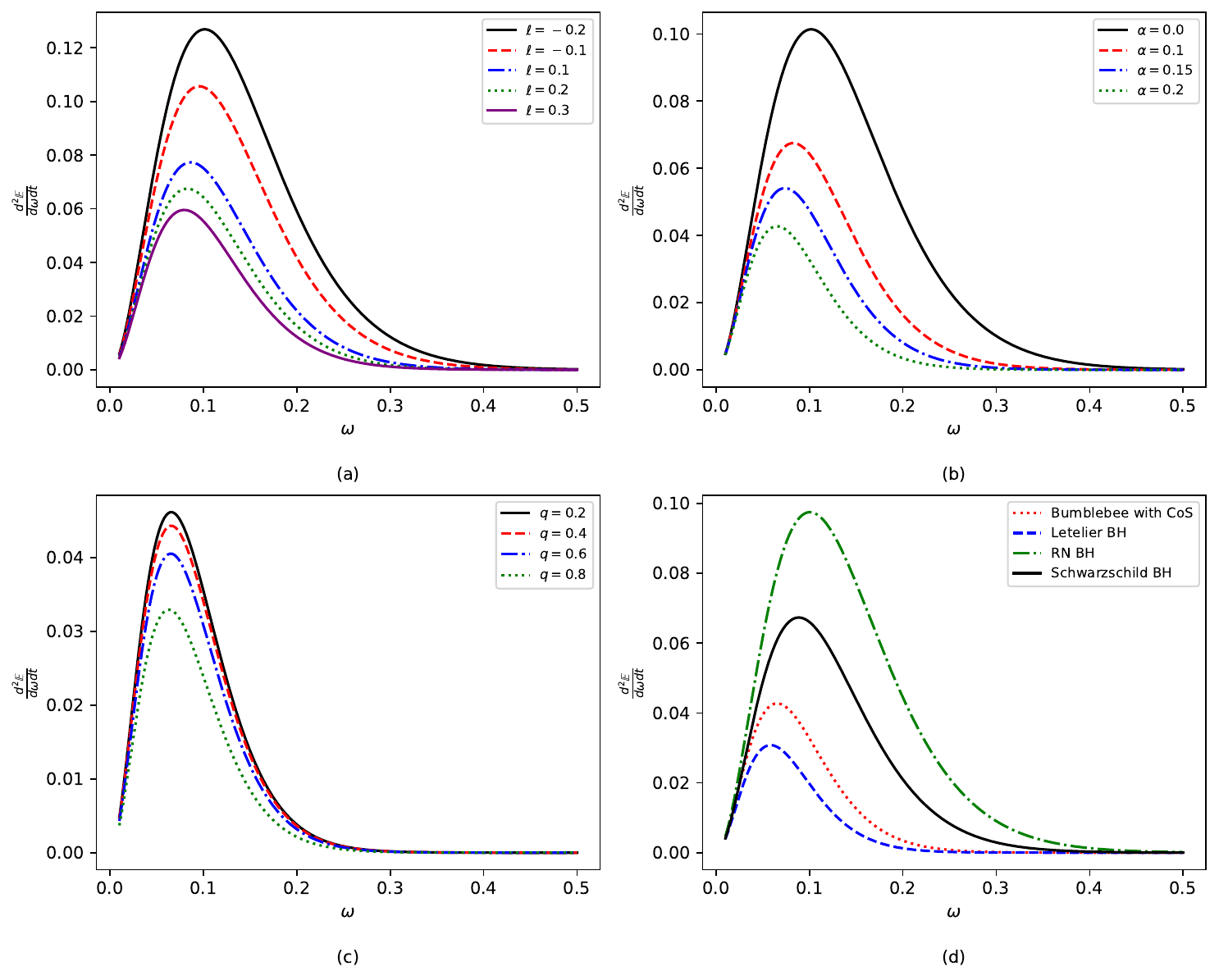}}
	\caption{Variation of energy emission rate $\frac{d^2\mathbb{E}}{d\omega\,dt}$ as a function of frequency in the bumblebee black hole spacetime with a cloud of strings (CoS). Panel (a) shows the variation with respect to the LV parameter $\ell$, panel (b) illustrates the influence of the CoS parameter $\alpha$, panel (c) depicts the dependence on the charge parameter $q$, and panel (d) presents a comparison between the bumblebee black hole with CoS, Letelier black hole, Reissner--Nordstr\"om black hole, and Schwarzschild black hole.}
\label{fig5}
\end{figure*}

\section{Rate of Energy Emission}\label{sec:8}

Quantum field theory in curved spacetime predicts that black holes are not perfectly black objects. Instead, they emit a thermal flux from the near-horizon region with a temperature determined by the surface gravity. In the semiclassical picture, this phenomenon can be interpreted as particle-antiparticle pair production near the event horizon, where one particle escapes to infinity while the other falls into the black hole. This process, known as Hawking radiation, results in a gradual loss of mass and energy from the black hole~\cite{Javed2019}. For a distant observer, the total radiated power depends not only on the Hawking temperature but also on the transmission probability that the emitted quanta overcome the gravitational potential barrier separating the horizon from spatial infinity.

In the geometric-optics (high-frequency) limit, the absorption cross section oscillates around a constant value~\cite{Misner1973,Mashhoon1973,Wei2013,Kala:2025xnb}
\begin{equation}
    \sigma_{\rm lim} \approx \pi R_{\rm sh}^2,
    \label{ee1}
\end{equation}
where $R_{\rm sh}$ denotes the black hole shadow radius given in Eq.~\eqref{dd12}. The proportionality $\sigma_{\rm lim} \propto R_{\rm sh}^2$ has a clear physical interpretation: at sufficiently high energies, the dominant contribution to the absorption cross section arises from photons (and other massless particles) with impact parameter equal to the critical value $\beta_c$, i.e., those that asymptotically approach the photon sphere. Consequently, the shadow radius effectively characterizes the absorptive area of the black hole in the high-frequency regime, establishing a direct connection between its optical appearance and thermodynamic properties.

Within this geometric--optics approximation, the spectral energy emission rate is described by a greybody-corrected Planck distribution~\cite{Wei2013},
\begin{equation}
    \frac{d^2\mathbb{E}}{d\omega\,dt}
    =
    \frac{2\pi^2\,\sigma_{\rm lim}}{e^{\omega/T}-1}\,\omega^3,
    \label{ee2}
\end{equation}
where $\omega$ represents the emitted frequency and $T$ is the Hawking temperature defined in Eq.~\eqref{bb4}.

Substituting Eqs.~\eqref{bb4} and~\eqref{dd12} into Eq.~\eqref{ee2}, the full emission rate for a static distant observer is given by
\begin{align}
\frac{d^2\mathbb{E}}{d\omega\,dt}=\frac{2\pi^3\,r_s^2 \omega^3\,(1-\alpha)}{1-\alpha-\frac{2 M}{r_s}+\lambda\frac{q^2}{r^2_s}}\,\left[\exp\left\{\frac{2\pi \omega r_h \sqrt{1+\ell}}{\frac{M}{r_h}-\lambda\frac{q^2}{r_h^2}}\right\}-1\right]^{-1}.\label{ee3}
\end{align}
Here $r_h$ is the event horizon and $r_s$ is the photon sphere radius given in Eq.~(\ref{dd8}).

The energy emission rate~(EER) is thus influenced by various geometric parameters of the model: the LSB parameter $\ell$, the string cloud parameter $\alpha$, the electric charge $q$, the black hole mass $M$, and the frequency $\omega$ for a given horizon and photon sphere radii $r_h$  and $r_s$, respectively. 

In Fig.~\ref{fig5}, we present the EER as a function of the emitted frequency for a bumblebee black hole embedded in a CoS background. We analyze the influence of individual model parameters on the EER and compare the results with several well-known black hole spacetimes that emerge as limiting cases of the present solution. From panel (a), it is clearly observed that the EER attains its maximum for smaller values of the LV parameter $\ell$ and gradually decreases as $\ell$ increases. This behavior indicates that Lorentz symmetry breaking suppresses Hawking radiation by modifying the effective horizon structure and the associated surface gravity. Furthermore, the EER is also found to decrease with an increase in the CoS parameter $\alpha$, reaching its maximum value at $\alpha=0$. Physically, the presence of the string cloud contributes an additional gravitational potential that reduces the effective temperature of the black hole, thereby lowering the emission rate. In addition, the EER decreases with increasing charge parameter $q$. However, the variation in EER with respect to $q$ is comparatively weaker than that induced by $\ell$ and $\alpha$, implying that the Lorentz-violating and CoS effects dominate over the electromagnetic contribution in shaping the radiation spectrum for the chosen parameter range. Finally, as shown in panel (d), the Reissner--Nordstr\"om black hole exhibits the maximum EER, whereas the Letelier black hole shows the minimum emission rate. These results suggest that, relative to the Schwarzschild geometry, the presence of electric charge enhances the EER due to an increase in the effective photon sphere radius, while the inclusion of the bumblebee field and the cloud of strings background suppresses the energy emission rate by reducing the surface gravity and modifying the spacetime curvature near the horizon. The analysis collectively demonstrates that Lorentz symmetry breaking and string cloud environments act as damping mechanisms for Hawking radiation, whereas the electric charge tends to enhance the energy emission rate, leading to distinct observational signatures for different black hole models.

\begin{figure*}[tbhp]
	\centerline{
		\includegraphics[width=80mm,height=70mm]{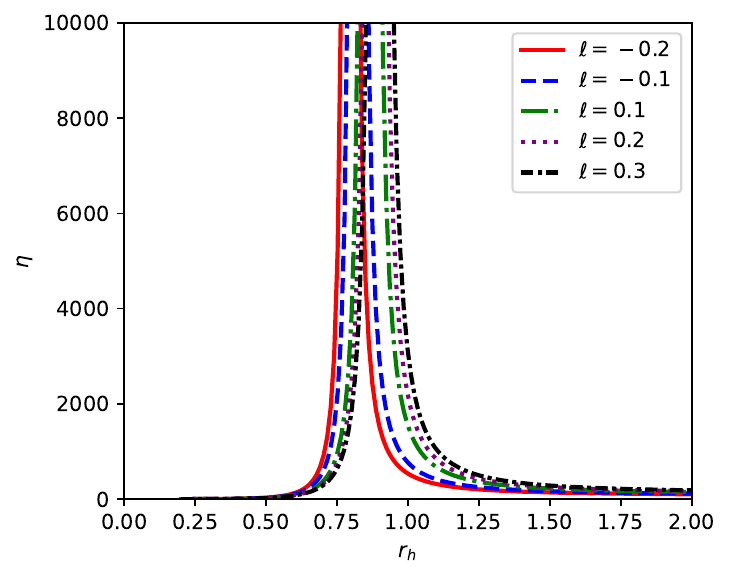}\qquad
        \includegraphics[width=80mm,height=70mm]{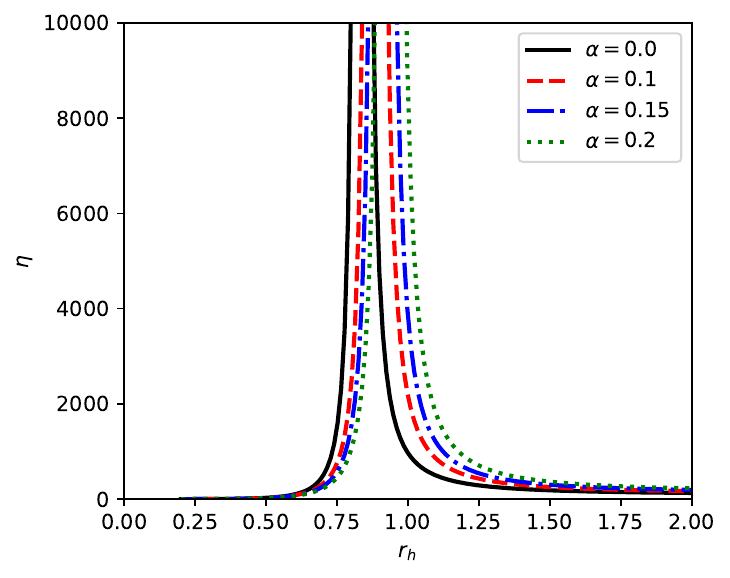}}(a) $\alpha=0.1$ \hspace{5cm} (b) $\ell=0.2$
	\caption{The variation sparsity as a function of $r_{h}$; for different values of LV parameter ($\ell$) and CoS parameter ($\alpha$). Here we fix, $M=1$, $q=0.8$, $\ell=0.2$ and $\alpha=0.1$ where applicable.}
\label{fig1SPARSITY}
\end{figure*}

\begin{figure*}[tbhp]
	\centerline{
		\includegraphics[width=180mm,height=70mm]{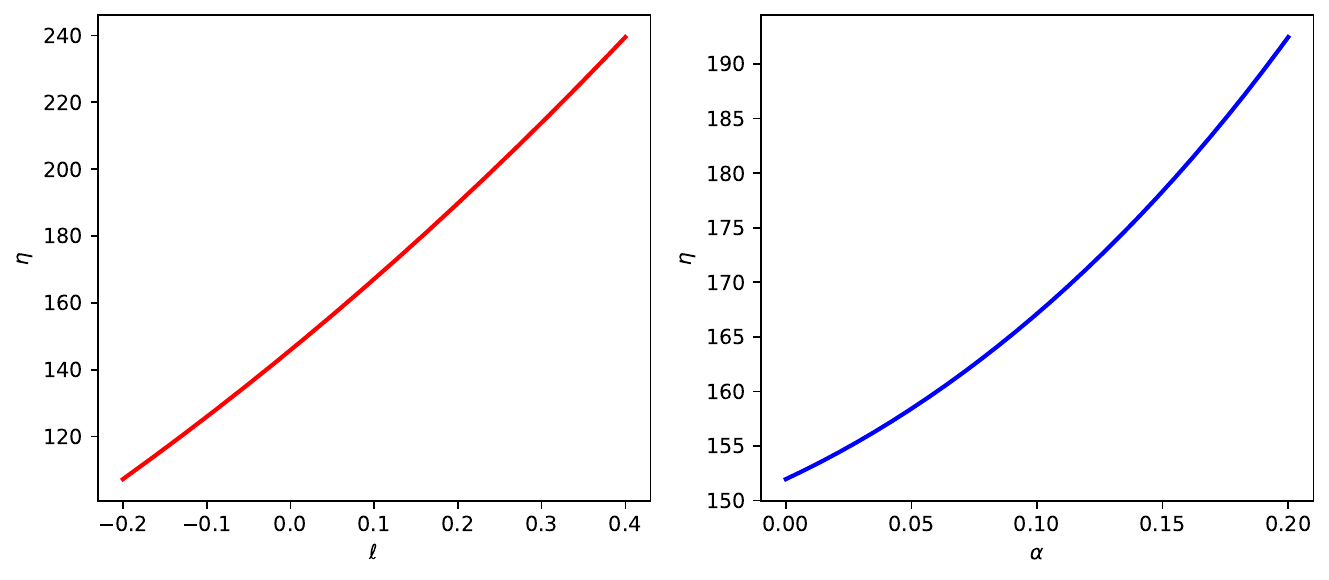}
        }(a)  \hspace{10cm} (b) 
	\caption{The variation sparsity as a function of of LV parameter ($\ell$) and CoS parameter ($\alpha$). Here we fix, $M=1$ and $q=0.8$.}
\label{fig2SPARSITY}
\end{figure*}

\section{Sparsity of Hawking radiation}\label{sec:9}

In this section, we quantify the sparsity of Hawking radiation for our black hole solution. Although a black hole radiates thermally at a temperature determined by its surface gravity, the Hawking emission is temporally discrete, consisting of well-separated quanta rather than a continuous flux. Sparsity is conventionally characterized by comparing the square of the thermal wavelength \(\lambda_t=2\pi/T\) with the effective emission area \(\mathcal{A}_{\rm eff}\). Following the definition introduced in \cite{PhysRevD.13.198, Gray_2016}, the dimensionless sparsity parameter is given by
\begin{equation}\label{defSpars}
    \eta =\frac{\mathcal{C}}{\Tilde{g} }\left(\frac{\lambda_t^2}{\mathcal{A}_{\rm eff}}\right),
\end{equation}
where \(\mathcal{C}\) is a dimensionless constant, \(\tilde g\) the spin degeneracy of the emitted quanta and  \(\mathcal{A}_{\rm eff}=\frac{27}{4}\,A_{\rm BH}=27\pi r_h^2\), where the horizon area in our case is $A_{\rm BH}=4 \pi r_h^2$. 

For our bumblebee gravity sourced by a cloud of strings of charged black hole solution, the modified surface gravity and horizon geometry influence both the Hawking temperature and the effective emission area. Consequently, the sparsity parameter exhibits a nontrivial dependence on the parameters $(\alpha,\,\ell,\,q)$, directly affecting the discreteness and observational signatures of the evaporation process. 

Substituting the temperature \eqref{bb9} into the sparsity definition \eqref{defSpars} yields the following closed form
\begin{equation}\label{sspp}
\eta=\frac{64 \pi ^3 (1+\ell) r_h^4}{27 \left[\lambda q^2-(1-\alpha) r_h^2 \right]^2},\end{equation}
where the horizon $r_h$ is in (\ref{horizon}).

Fig.~\ref{fig1SPARSITY} shows how the sparsity parameter eta of Hawking radiation changes with the horizon radius $r_h$ for various values of the $\ell$ and $\alpha$. In panel (a), the parameter $\alpha$ is kept fixed while the value of $\ell$ is varied. It is observed that the sparsity parameter $\eta$ shows a sharp divergence near a critical value of $r_h$, which corresponds to the regime where the Hawking temperature approaches zero. Away from this region, $\eta$ decreases rapidly as $r_h$ increases, indicating that the radiation becomes less sparse around larger black holes. Increasing the LV parameter $\ell$ shifts the peak of the sparsity curve and generally raises the magnitude of $\eta$, indicating that Lorentz-violating effects increase the sparsity of the Hawking radiation. In panel (b), the value of $\ell$ remains constant while $\alpha$ changes. A similar qualitative behavior is observed in which larger values of the CoS parameter alpha lead to an increase in the sparsity parameter and a slight shift in the divergence point. This shows that the emission characteristics of the Hawking radiation are influenced by both the LV parameter and the cloud of strings parameter.\\
Fig.~\ref{fig2SPARSITY} shows the dependence of the sparsity parameter $\eta$ directly on the parameters $\ell$ and $\alpha$. Panel (a) demonstrates that $\eta$ increases monotonically with the LV parameter $\ell$, indicating that stronger Lorentz-violating effects lead to more sparse Hawking radiation. Similarly, panel (b) shows that the sparsity parameter also increases with CoS parameter $\alpha$. These results suggest that both the LV parameter and the string cloud background tend to suppress the emission rate of Hawking quanta, thereby increasing the sparsity of the radiation.
\begin{table}[h]
\centering
\caption{Numerical values of the sparsity parameter $\eta$ for different values of the Lorentz-violating parameter $\ell$ and the cloud of strings parameter $\alpha$ for $M=1$ and $q=0.8$.}
\begin{tabular}{c|cccc}
\hline
$\alpha$ & $\ell=-0.1$ & $\ell=0$ & $\ell=0.1$ & $\ell=0.2$ \\
\hline
0.0 & 111.25 & 130.66 & 151.97 & 175.37 \\
0.1 & 125.93 & 145.86 & 167.12 & 189.77 \\
0.2 & 148.03 & 169.74 & 192.46 & 216.20 \\
\hline
\end{tabular}
\label{table:sparsity}
\end{table}

Table~\ref{table:sparsity} shows the numerical values of the sparsity parameter $\eta$ for different values of the Lv parameter $(\ell)$ and the CoS parameter $(\alpha)$. It is observed that $\eta \gg 1$ in all cases, indicating that the Hawking radiation is highly sparse. Furthermore, the sparsity parameter increases with increasing values of both $\ell$ and $\alpha$, suggesting that Lorentz symmetry breaking and the presence of a CoS enhance the discreteness of the Hawking emission.

\section{Conclusions}\label{sec:10}

In theories where Lorentz symmetry in gravity is spontaneously broken, a nonminimally coupled bumblebee vector field acquires a nonzero vacuum expectation value, leading to modifications of standard GR. In this work, we investigated exact solutions describing static and spherically symmetric charged black holes surrounded by a cloud of strings within the framework of bumblebee gravity.

We first analyzed the thermodynamic properties of the considered black hole solution, including the Hawking temperature, Helmholtz free energy, specific heat, and Gibbs free energy. We examined how the Lorentz-violating parameter and the presence of a string cloud jointly modified these thermodynamic quantities in comparison with the standard charged black hole case, namely the Reissner-Nordström black hole. To illustrate these effects, we generated several plots showing how the thermodynamic quantities varied with the Lorentz-violating parameter and the string cloud parameter.

We then studied the optical properties of the spacetime by analyzing the photon sphere, the resulting black hole shadow, and the effective radial force experienced by photon particles, thereby identifying possible observational signatures of Lorentz symmetry violation and the presence of a string cloud. In addition, we investigated the bending of light and derived an expression for the deflection angle in the weak-field limit.

Furthermore, we explored the influence of the Lorentz-violating and string cloud parameters on classical gravitational tests within the Solar System, particularly the advance of perihelion precession. Using a perturbative method, we derived the corresponding perihelion shift and explicitly demonstrated the individual contributions of the Lorentz-violating parameter and the string cloud.

Finally, we computed the energy emission rate and sparsity of Hawking radiation of the black hole and demonstrated how the Lorentz-violating and string cloud parameters modified the energy rate compared with the standard charged black hole case. Moreover, we observe that the dimensionless sparsity parameter $\eta$ is more than that of the standard RN-black hole case due to the presence of Lorentz-violating and cloud of strings. Overall, our analysis provided a comprehensive investigation of the interplay between Lorentz violation, black hole physics, and a cloud of strings, offering a potential framework for probing new physics beyond GR.

\footnotesize

\section*{Acknowledgments}

F.A. acknowledges the Inter University Center for Astronomy and Astrophysics (IUCAA), Pune, India for granting visiting associateship. The author, SK, sincerely acknowledges IMSc for providing exceptional research facilities and a conducive environment that facilitated his work as an Institute Postdoctoral Fellow.

\section*{Data Availability Statement}
There are no new data associated with this article [Authors comment: No data were generated in this article]..

\section*{Code/Software}

No code/software were developed in this article [Authors comment: No code/software were developed in this article].

\bibliographystyle{apsrev4-2}
\bibliography{main}

\end{document}